\documentclass[journal]{IEEEtran}
\pdfoutput=1
\usepackage{times,amsmath,epsfig}
\usepackage{epstopdf}
\usepackage{hyperref}
\usepackage{cite}
\usepackage{color}
\usepackage{subfigure}
\usepackage{comment}
\usepackage{stfloats}
\usepackage{multirow}
\usepackage{threeparttable}

\newcommand{\tabincell}[2]{\begin{tabular}{@{}#1@{}}#2\end{tabular}}
\usepackage{diagbox}
\title{Perceptual Quality Assessment of Immersive Images Considering Peripheral Vision Impact}
\author{Peiyao Guo, Qiu Shen, Zhan Ma,  David J. Brady, and Yao Wang}
\begin{document}
\maketitle



\begin{abstract}
Conventional images/videos are often rendered within the central vision
area of the human visual system (HVS) with uniform quality.
Recent virtual reality (VR) device with head mounted display (HMD)
extends the field of view (FoV) significantly to include both central
and peripheral vision areas. It exhibits the unequal image quality sensation among these areas
because of the non-uniform distribution of photoreceptors on our retina.
We propose to study the sensation impact
on the image subjective quality with respect to the eccentric angle $\theta$ across different
vision areas. Often times, image quality is controlled by the quantization stepsize $q$ and spatial resolution $s$,
separately and jointly. Therefore, the sensation impact can be understood by exploring the $q$ and/or $s$
in terms of the $\theta$,  resulting in self-adaptive analytical models that
have shown quite impressive accuracy through  independent cross
validations.  These models can further be applied to give
different quality weights at different regions,
so as to significantly reduce the transmission data size but without subjective quality loss. As demonstrated
in a gigapixel imaging system, we have shown that the image rendering can be speed up about 10$\times$ with the model
guided unequal quality scales, in comparison to the the legacy scheme with uniform quality scales everywhere.
\end{abstract}


\section{Introduction}
\begin{figure*}[t]
	\centering
	\subfigure[]{\includegraphics[scale=0.155]{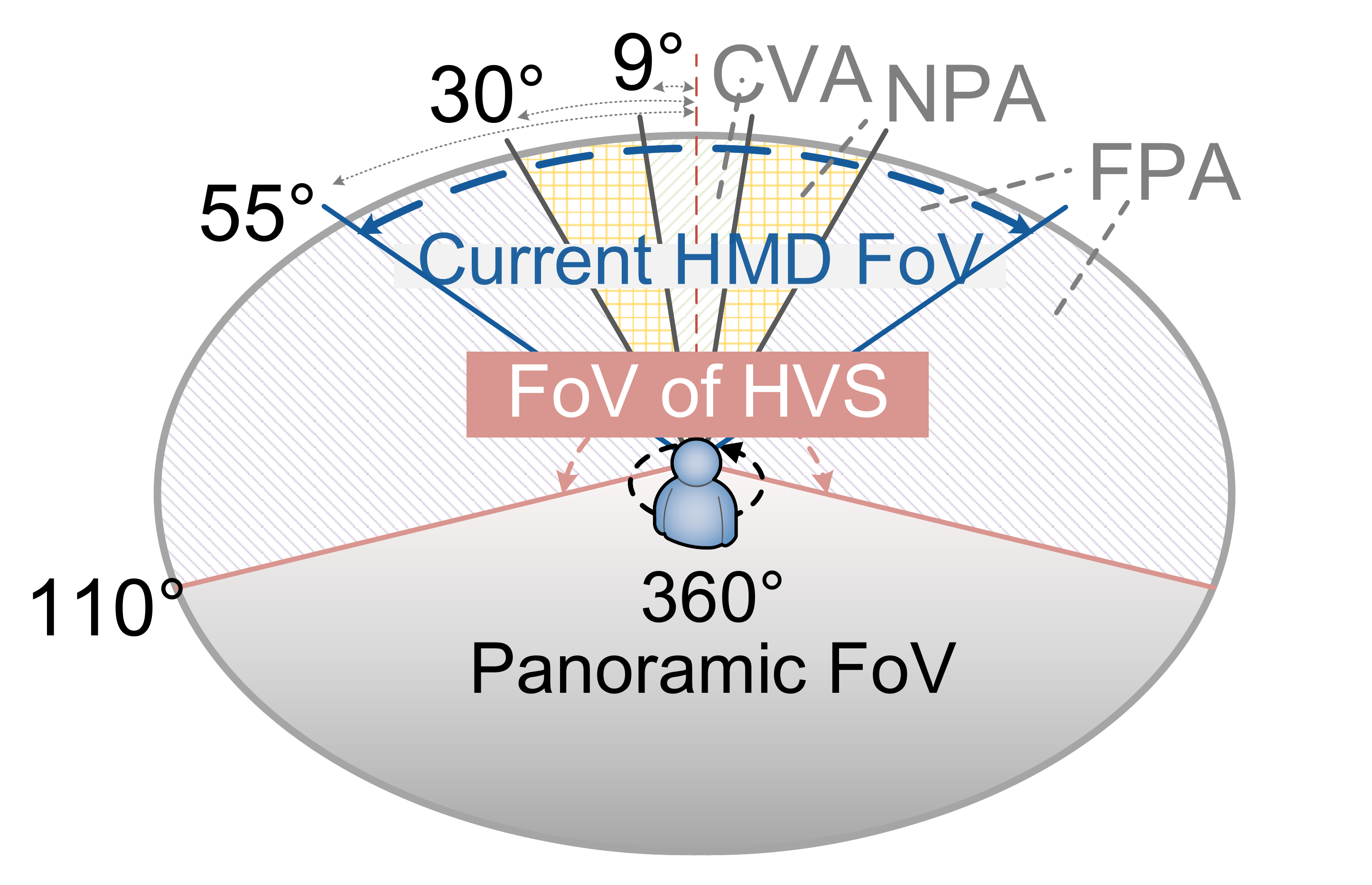} \label{sfig:2D_FoV}}
	\subfigure[]{\includegraphics[scale=0.5]{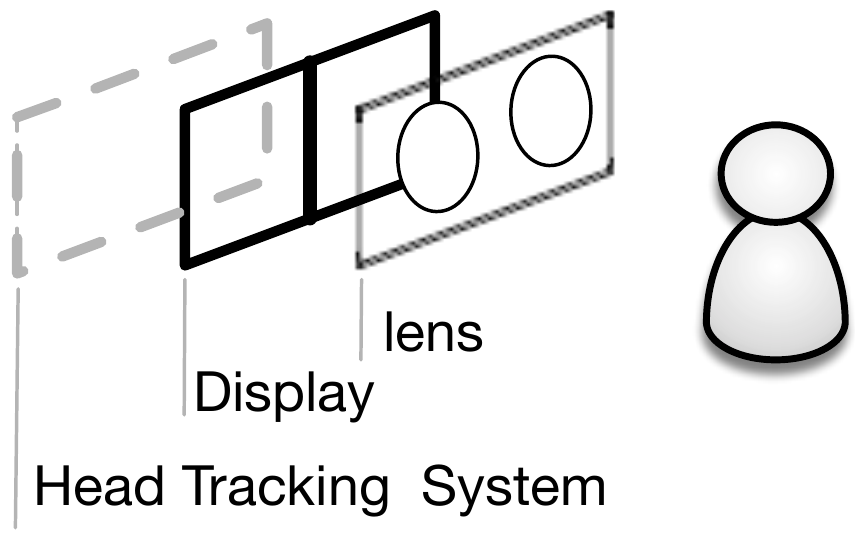}\label{sfig:hmd}}
	\subfigure[]{\includegraphics[scale=0.08]{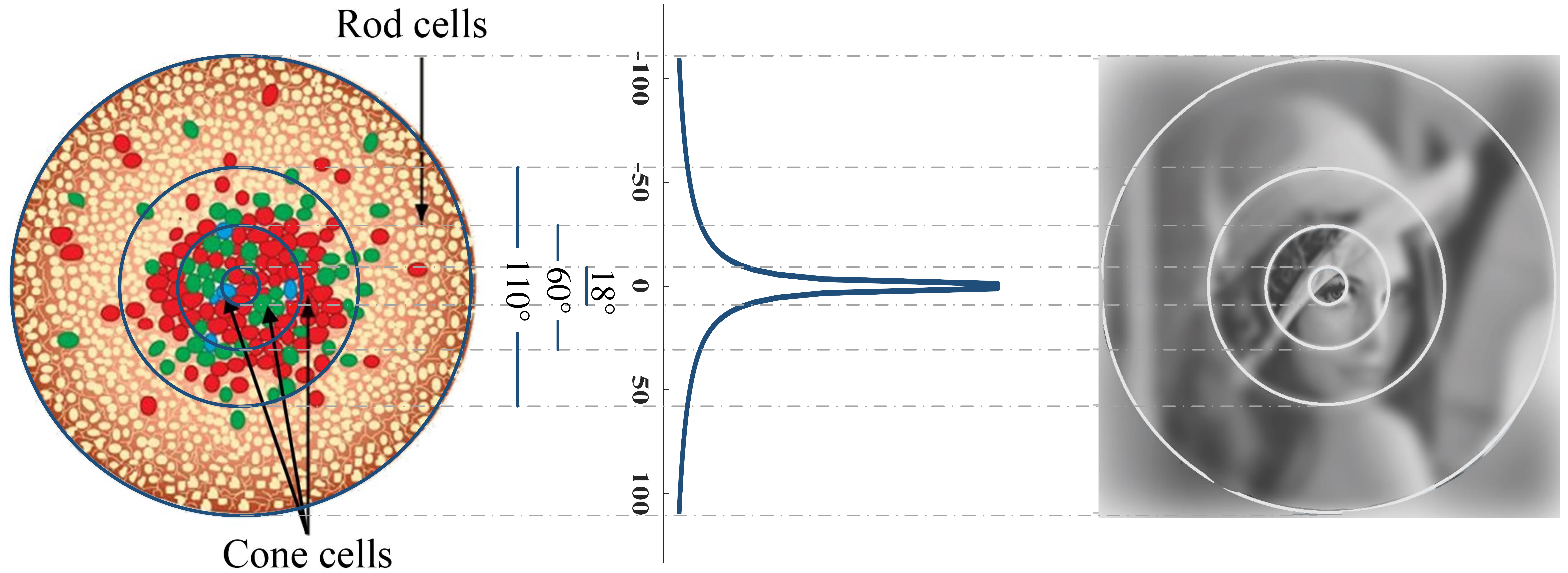}\label{fig:density of photoreceptor}}
	\caption{Illustration of the panoramic FoV exhibited by the head mounted display as well as the characteristics of the vision areas: (a)  central vision area (CVA), near peripheral area (NPA) and far peripheral area (FPA) of current FoV for a panoramic scene (b) an example of head mounted display for a typical virtual reality device  (c) the distribution of photoreceptors on the human retina and corresponding non-uniform visual perception}
	\label{fig:panoFoV}
\end{figure*}

We see the world through our eyes, with the binocular field of view (FoV) about 220$^\circ$ horizontally\cite{FoV_wiki}. {We follow the convention
to represent the range of FoV using the horizontal viewing range throughout this work unless we point out otherwise}.
Our FoV mainly includes three distinct regions as shown in Fig.~\ref{sfig:2D_FoV}, i.e.,
central vision (or macular) area (CVA) with one-side 9$^\circ$ eccentrically, near
peripheral area (NPA) with one-side 30$^\circ$ eccentrically and far
peripheral area (FPA) covering the rest region.

Usually, conventional video frames or images are
rendered within a very limited range in our current FoV (i.e., typically around 18$^\circ$ by calculating the distance
between the user and the display screen) on the flat display in the front,
such as the TV panel, mobile screen, etc.  Such viewing range overlaps with the {\it central vision} area of our human visual system (HVS).
It is far less than our visible FoV in reality.

Panoramic and immersive images/videos are becoming quite popular,
because of the recent introduction of a variety of powerful
virtual reality (VR) devices, such as the HTC Vive, Oculus
Rift, and the Samsung Gear VR. The basic elements of these
systems are a binocular head mounted display (HMD), with
head tracking hardware as shown in Fig.~\ref{sfig:hmd}. The experience
of viewing 2D or 3D immersive images in this manner can be
stunning, especially given a very wide panoramic FoV (i.e., binocular
110$^\circ$ of HTC Vive HMD that is much larger than the 18$^\circ$ of our central vision),
within which users can navigate and interact with their
virtual environment. The sensation of a vividly virtualized reality
can be dramatic compared to traditional viewing of images on
fixed display screens having very limited FoVs.

A topic of interest in this context is the perceived quality
of immersive VR images.
Conventionally, we have mainly studied the uniform subjective quality of the image and video by simply assuming that
they are rendered in the central vision with the highest sensation.
We name few of them here, i.e.,
the structural similarity (SSIM)~\cite{SSIM}, the just-noticeable-distortion (JND)~\cite{JND_imageCoder},
the quality model considering the spatial, temporal and amplitude resolutions of the compressed video (Q-STAR)~\cite{pv_mobileQSTAR,Ma_RQModel}, etc.
As seen, recent VR devices
extend the viewing range of the displayed content significantly (e.g., 110$^\circ$ of HTC Vive, or 90$^\circ$ of Samsung GearVR).
Although it is still less than our binocular 220$^\circ$ FoV, it already includes both the
central and peripheral vision areas.  However, to the best of our knowledge, we have not seen a systematic study
discussing the peripheral vision impact on the overall perceptual quality of immersive images. Such study would be of
great value in
the development of objective quality assessment predictors for
immersive images and videos, for analyzing the perceptual
impact of wireless transmission on immersive images and
videos, and so on.




Our HVS exhibits very different perceptual sensation in different areas according to the existing
research works~\cite{jov_peri_vision_review} on vision and neuroscience. For instance,
macular (central vision) area as aforementioned, which is about 9$^\circ$ eccentrically (i.e., from the center of our retina), often requires ultra high resolution and high fidelity, while visual perception would have very sharp degradation in
our periphery. This is because of the highly non-uniform distribution
of photoreceptors on the human retina~\cite{curcio1990human}, as shown in Fig.~\ref{fig:density of photoreceptor}.
Therefore, we could leverage this biological characteristics to distribute non-uniform image quality scales in different areas.
More specifically, high quality representation of the content is used in corresponding central vision area, but
reduced quality copies in peripheral areas. Naturally speaking, for a compressed image, a high quality version
normally demands more  bits for representation, but less bits consumption for reduced quality one~\cite{R-STAR}. Hence, such a non-uniform quality distribution in central and peripheral vision would lead to bandwidth consumption reduction,
but this requires a quantitative model that could explicitly express the peripheral vision impact on image perceptual quality.
This model could be used to give appropriate compression factors that save the bits consumption, but without incurring any
perceptual degradation.

Toward this goal, we propose to measure the image quality degradation in peripheral areas with respect to the
degree of eccentricity ($\theta$) from the center of the retina. {Here, we use the horizontal
eccentricity to simplify the illustration of the 2-D viewing range of a FoV.}  Note that the quality of a compressed image
is mainly determined by the quantization stepsize $q$\footnote{Images are  encoded with
the quantization stepsize $q$ via H.264/AVC intra codec.
$q = {2^{\frac{{{\rm QP} - 4}}{6}}}$~\cite{ma2005study}, in which QP represents the quantization parameter.  We use QP and $q$ interchangeably throughout this work.} and spatial resolution $s$~\cite{Ma_RQModel,VR_spatial_IC3D,pv_mobileQSTAR}.
Thus the problem can be rephrased to devise appropriate mathematical forms to describe the
variations of $q$ and $s$, with respect to the $\theta$, separately and jointly, referred to as the $q$-impact, $s$-impact and joint $q$-$s$-impact,
respectively.

We have invited
hundreds of subjects with normal vision to participate the subjective quality assessment
for peripheral vision impact study. Both $q$ and $s$ can
be well modeled using a generalized parametric  Gaussian model in terms of the $\theta$ separately.
We have further discovered that  $q$-impact is independent of the $s$ through the joint $q$-$s$-impact
exploration. By connecting with the perceptual quality models assuming the uniform image quality in our previous works~\cite{VR_spatial_IC3D,Xiaokai_TIP}, we could finally reach a closed-form perceptual quality model for immersive image with the consideration of the peripheral vision impact.

Parameters are either fixed or can be estimated using the content features.
Proposed models are then utilized to generate images with non-uniform quality scales (using different
$q$ or $s$ in peripheral area) for independent cross-validations using another set of data.
Note that the cross-validation is performed to rate the same immersive image but different quality setups, i.e., one is with the conventional uniform
quality while the other one is with the unequal quality to the content regions in the central and peripheral areas.
It demonstrate the effective
results with both Pearson correlation (PCC) and Spearman Rank Correlation coefficients (SRCC) larger than 0.92.

We then devise this quality model for fast gigapixel image retrieval and rendering with the head mounted display,
given that the model driven unequal quality setup could
reduce the data size significantly without incurring the perceptual quality degradation.
Our tests have shown that with the unequal quality setup, image retrieval time is about 10$\times$ reduction. This would be
very useful for the gigapixel image navigation application.

\begin{figure*}[t]
\centering
\subfigure[Attic]{\includegraphics[scale=0.059]{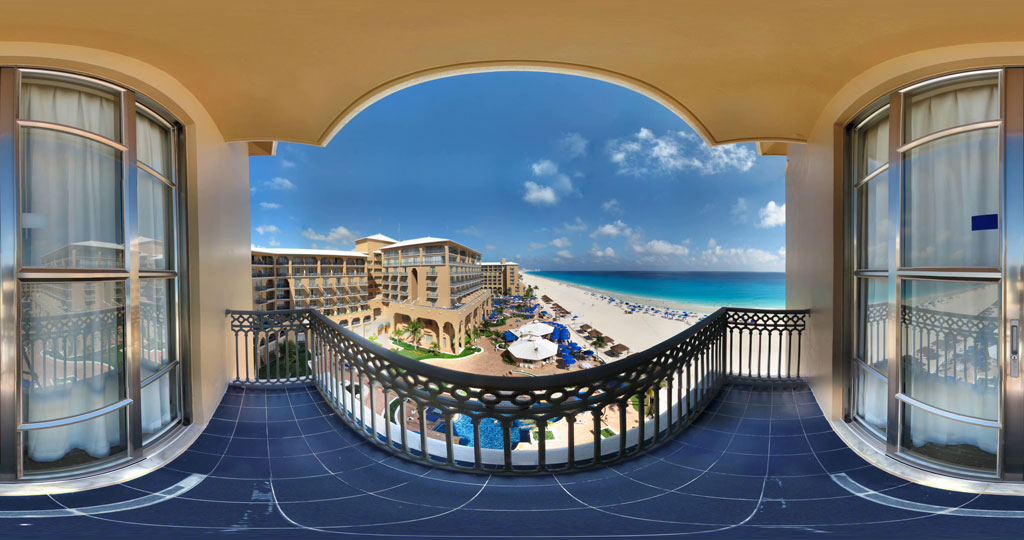}}
\subfigure[Temple]{\includegraphics[scale=0.059]{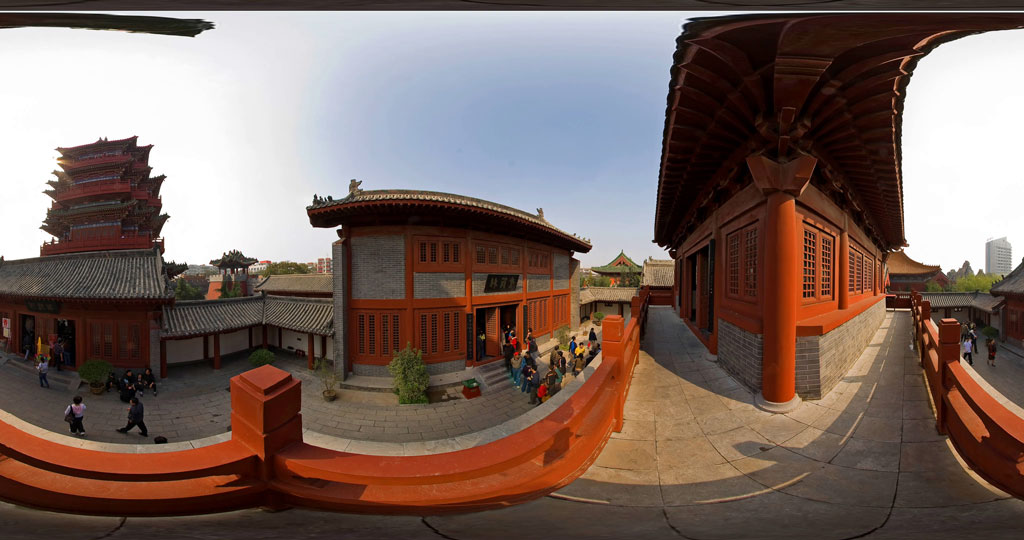}}
\subfigure[Ship]{\includegraphics[scale=0.059]{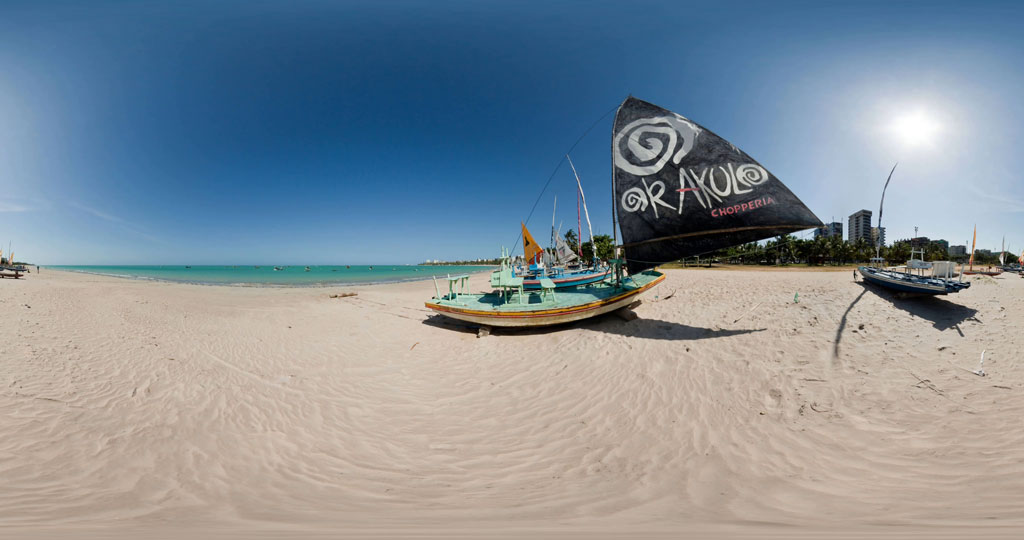}}
\subfigure[Train]{\includegraphics[scale=0.059]{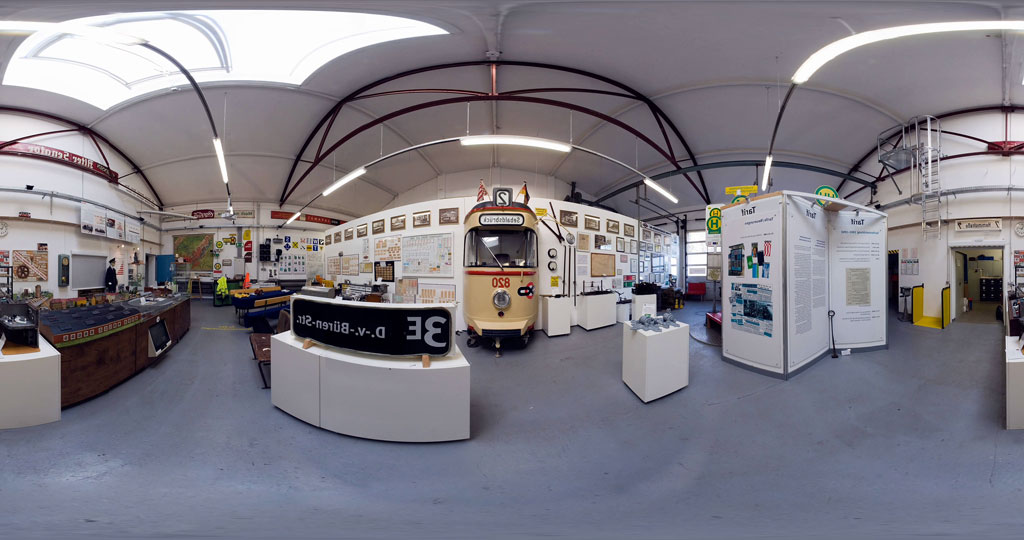}}
\subfigure[Beach]{\includegraphics[scale=0.059]{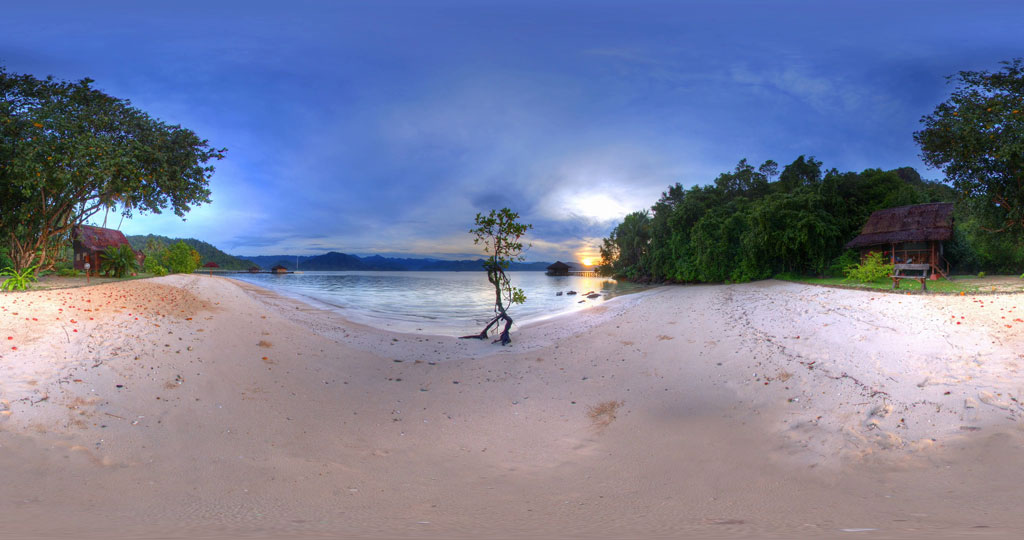}}
\subfigure[Sculpture]{\includegraphics[scale=0.059]{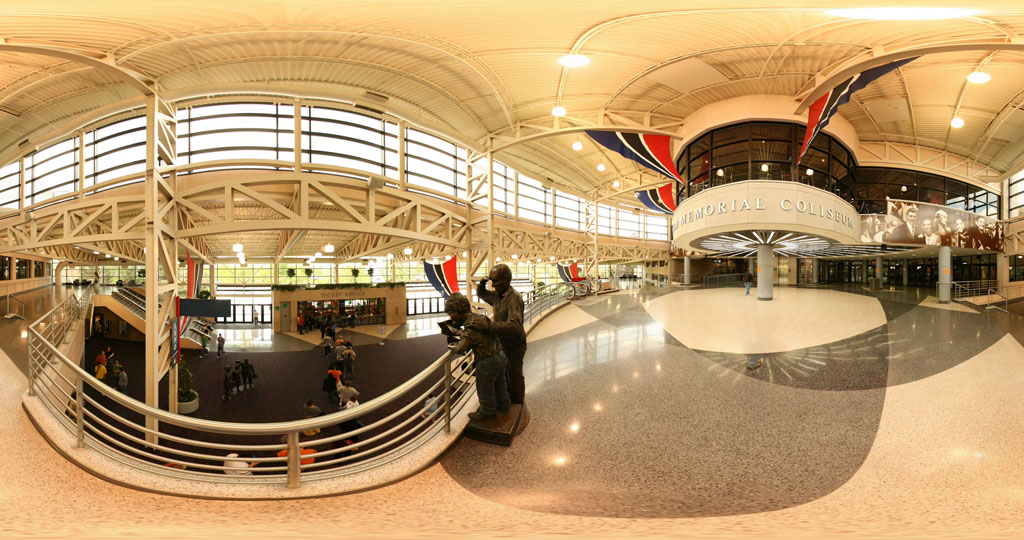}}
\subfigure[Football]{\includegraphics[scale=0.059]{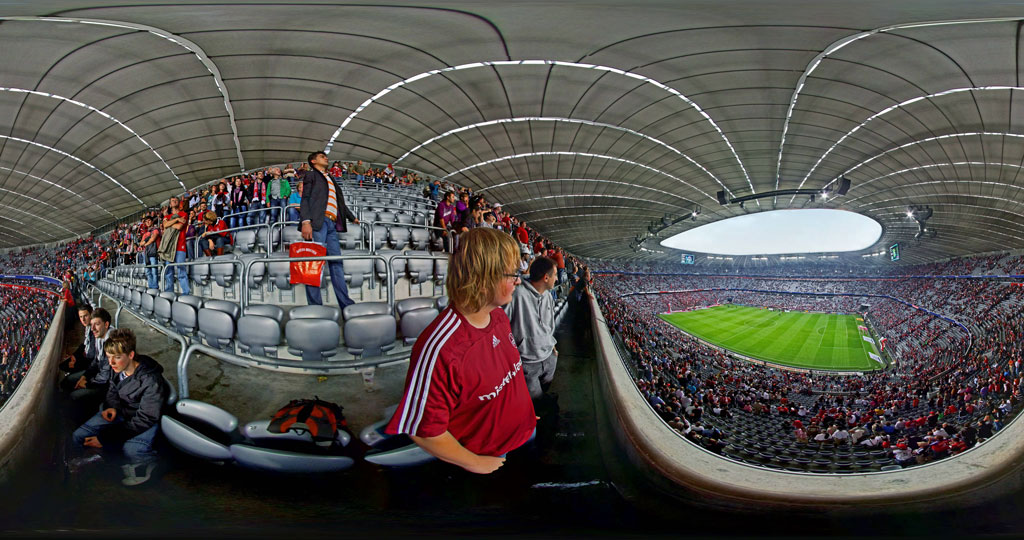}}
\subfigure[Desert]{\includegraphics[scale=0.059]{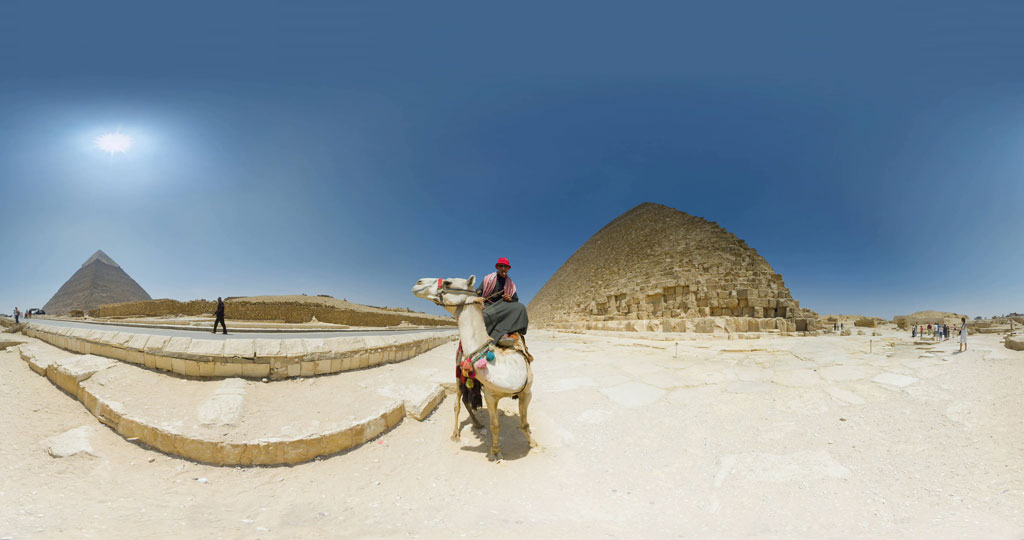}}
\caption{Examples of immersive images used for peripheral vision impact study.}
\label{fig:sourcedata_model_development}
\end{figure*}

The reminder parts of this work are organized as follows: Section~\ref{sec:related_work} first introduces relevant studies in the literature, and then Section~\ref{sec:peri_model} explains the details regarding how to measure the peripheral vision impact on an immersive image shown in HMD, and propose analytical models to quantify the $q$ and $s$ with respect to $\theta$ separately or jointly. The proposed peripheral vision model is cross-validated in Section~\ref{sec:cross-validation}. The proposed model is applied to guide the fast retrieval of the gigapixel images in Section~\ref{sec:application}. Finally, the conclusion is drawn in Section~\ref{sec:conclusion}.

\section{Related Work} \label{sec:related_work}

This work greatly appreciates those efforts devoted in the area of image and video quality assessments (IQA).
Thus, we review several of them that are commonly adopted in practice.

Following the distortion measurement
of the analog signal, mean square error (MSE) or similar peak signal-to-noise ratio (PSNR) is easily
extended to evaluate the distortion of the image or video, assuming the quality is directly related to the
pixel amplitude degradation (i.e., due to the noise or compression) and each pixel is equally important.

Our HVS often plays magic when viewing the image or video by selectively ignoring or emphasizing
certain regions (i.e., masking~\cite{constrast_masking,spatial_distortion,color_quality}). Particularly, users are very sensitive to the structural
distortion instead of perceiving the amplitude loss of a single pixel. Hence, structural similarity~\cite{SSIM} can be well captured by the mean, variance and co-variance of the original and distorted images.
All the computations are evaluated in
pixel domain as the MSE or PSNR.

National
Telecommunications and Information
Administration (NTIA) has published a generalized video quality metric (VQM)~\cite{VQM}
based on the extensive subjective studies performed in Video
Quality Experts Group (VQEG) Phase II Full
Reference Television tests. It suggested that the distortion of an image could be
represented by the weighted distortions of several content features (i.e., seven in total),
including spatial activities,
 edge distributions, chroma spread,  etc.  Here, we need to first calculate the features from the original
 and distorted image, and then derive the final distortion score in features' domain.

All aforementioned metrics belong to the full reference category that requires both the distorted
and original image sources. However, in many application scenarios, original sources are not available.
Instead, we could utilize some partial informations (such as wavelet transform coefficients~\cite{ZWang_RR_IQA},
divisive normalization based representation~\cite{DNT_jstsp09}, etc.)
extracted from the original image (but with much lower
data rate) to help the distortion evaluation.

Ultimately, we expect to measure the image quality blindly~\cite{xli_IQA,ZWang_NR_IQA}
without any reference. Most of them are developed based on the natural scene statistics (NSS)~\cite{NR_NSS}£¬
either in spatial domain~\cite{BRISQUE,NIQE}, or in transform domain (e.g., DCT~\cite{BLIINDS-2}), etc.
Machine learning~\cite{BLIINDS-2,DIIVINE} can be also introduced to further the performance.
All above non-reference quality assessment methods are mainly focusing on the image/video with fixed spatial/temporal
resolution. Intuitively, compressed video can be represented as a function of its spatial resolution (e.g., frame size).
temporal resolution (e.g., frame rate) and amplitude resolution (e.g., quantization induced pixel amplitude degradation)
for its perceptual quality~\cite{pv_mobileQSTAR,VQMTQ} and bit rate~\cite{R-STAR}. Together with the content adaptive
parameters, these models can be easily applied to do multi-dimensional optimization~\cite{STARopt}.

More recently, we have performed some preliminary studies to model the quality of immersive images rendered on the
head mounted displays~~\cite{VR_spatial_IC3D,Xiaokai_TIP} considering the impacts of the spatial resolution and quantization.
However, these works still follow the traditional methodology that an uniform quality (i.e., quantization or spatial resolution) is
applied for the entire immersive image without taking into the unequal fidelity sensation of our HVS between the central and peripheral vision areas.   In the coming paragraphs, we will discuss the perceptual quality modeling of the immersive image
considering the peripheral vision impact.

\section{Perceptual Quality Modeling of Immersive Image Considering the Peripheral Vision Impact} \label{sec:peri_model}
It is known that human eye has different spatial resolution distinguishability
 between central and peripheral vision area because of the highly non-uniform distribution
of photoreceptors on the human retina~\cite{curcio1990human}, shown in Fig.~\ref{fig:density of photoreceptor}. Tyler~\cite{tyler1997analysis} has proposed a power function to quantify the density of cones (e.g., measured by
the number of cones per mm$^2$)
 from 1$^\circ$ to 20$^\circ$ eccentrically, i.e.,
\begin{equation}
	\rho(\theta)= 50000 \cdot {\left( {\frac{\theta }{{300}}} \right)^{-\frac{{ 2}}{3}}},  \mbox{~~~}  \theta\in[1^{\circ}, 20^{\circ}],
\end{equation}
while $\rho(\theta)$ is linearly decreased until 4000 cones/mm$^2$ with $\theta > 20^{\circ}$.  Mathematically, Tyler's $\rho(\theta)$ can be
seen as an approximation of the generalized Gaussian distribution.

Intuitively, users have sensitive perception in the area with the  higher density  of  cones, but may not tell the difference of image degradation in the area with
less cones.  We believe that the ability to distinguish the image perceptual quality variation should follow the density distribution of the photoreceptors $\rho(\theta)$ in our retina.
As the image quality can be represented by its signal fidelity (that  is often controlled using $q$ or $s$)~\cite{pv_mobileQSTAR}, the overall problem is to model the $q$ or $s$ with respect to the degree of eccentricity $\theta$ without noticing the image quality degradation perceptually.
Towards this goal,
we have carefully designed the subjective tests to collect users' opinion scores, i.e. mean opinion score (MOS)  so as to develop the analytical models.

\subsection{Subjective Experimental Setup} \label{ssec:subjective_assessment}

\subsubsection{Testing Platform}
\begin{figure}[b]
    \centering
    \includegraphics[scale = 0.49]{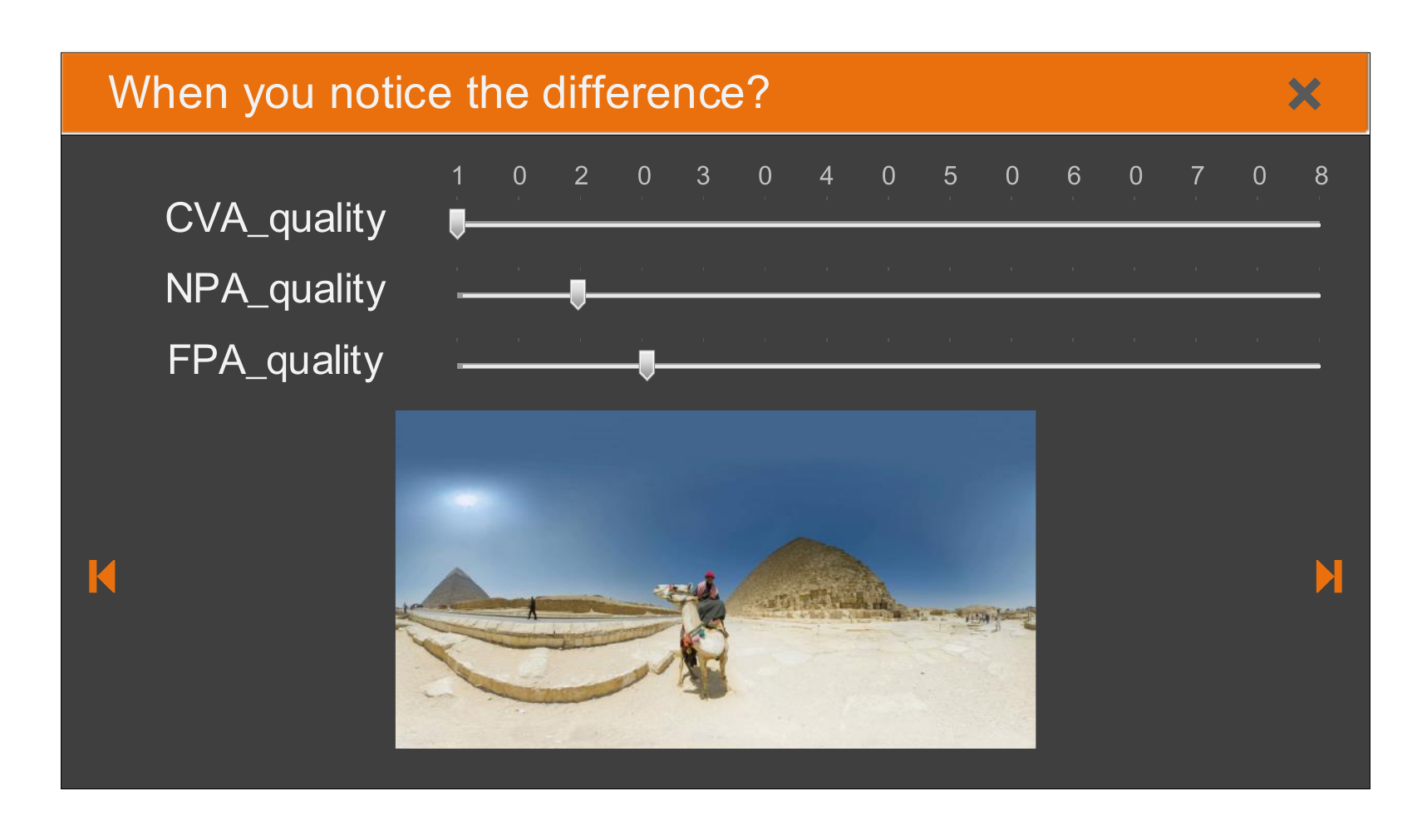}
    \caption{Screenshot of the interactive control console interface. This interface is installed in the HTC Vive system connected computer and a technician is required to operate the quality factor adaptation.}
    \label{fig:gui}
\end{figure}

We choose the HTC Vive system ~\cite{HTC_vive} with its associated HMD  to perform the subjective quality assessments. This is mainly because the HTC  Vive platform offers the state-of-the-art user experience when enjoying the immersive content. It gives the user the freedom to navigate  and interact with the content inside the virtualized environment.  The HTC Vive HMD provides the binocular 110$^\circ$ FoV at $2160\times1200$ spatial resolution refreshed at 90Hz (or FPS).  To accurately control the $q$ and $s$ when performing the assessments, we have implemented
an interactive UI, as shown in Fig.~\ref{fig:gui}.  This control panel is displayed on the screen of a powerful computer paired with the HTC Vive system for rendering.  A technician will manually adapt the quality factors to display different image pairs, when subject wear the HMD and perform the rating process.

\subsubsection{Test Sequence Pool}
\begin{figure}[t]
    \centering
    \includegraphics[scale =0.8]{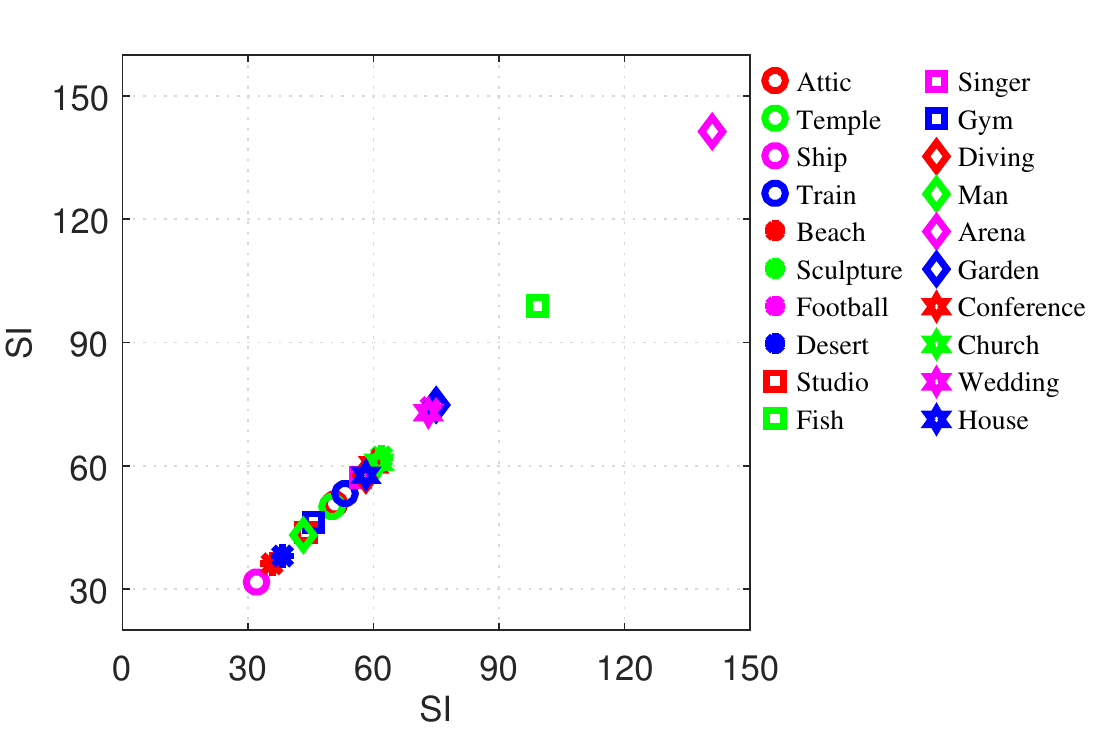}
    \caption{The spatial  information indices of the test sequences.  }
    \label{fig:SI}
\end{figure}

Eight immersive images, i.e.,\emph{ Attic, Temple, Ship, Train, Beach, Sculpture, Football, Desert}, from the SUN360 database~\cite{xiao2012recognizing} downsampled to the spatial resolution at 4096$\times$2160 are chosen as our test materials, as shown in Fig.~\ref{fig:sourcedata_model_development}. Another twelve images are used for cross-validation shown in Fig.~\ref{fig:sourcedata}.
In total, we have randomly selected twenty images with their  spatial information index (SI)~\cite{yu2013image} presented in Fig.~\ref{fig:SI}. As seen, these test samples cover a wide range of the content characteristics, reprensenting typical scenarios of immersive applications.
Besides, each test image contains meaningful saliency area within user's FoV as rendered in the HMD.

 To ensure the derived model generally applicable, every image are prepared with multiple versions to cover the sufficient quality scales, via different combinations of the $s$ and/or $q$. As shown in Table~\ref{tab:quality_paras}, we have performed three independent tests to study the separable $q$-impact, $s$-impact as well as the joint $q$-$s$-impact. More specifically, for evaluating the independent $q$-impact, we have enforced the spatial resolution at its native resolution, ie., $s$ = $s_{\max}$ = $4096\times2160$, but applied ten different QPs (or $q$); Meanwhile, we have adapted $s$ with eight distinct levels for each raw image (i.e., QP = 0) to study the $s$-impact. For joint $q$-$s$-impact,  we still use ten different QPs, but with only four distinct spatial resolutions. One reason is to reduce the size of the test samples for each participants. This is because the subjects feels uncomfortable and tired after a very long rating process. Normally, we would like to keep the test duration for each subject less than 30 minutes~\cite{Xiaokai_TIP}.  In total, we recruit
175 students (aging from 18 to 30), including 101 male and 74 female, from different majors in Nanjing University to participate this assessment. All viewers have normal vision (or after correction) and color perception. About 90\% of viewers are naive with video processing, subjective assessment or virtual reality.



 \begin{table}[b]
	\centering
	\caption{Quality Control Parameters Used in Subjective Assessments }
	\label{tab:quality_paras}
	\begin{tabular}{|c|c|c|}
		\hline
		& QP & $s$\\
		\hline
		Test \#1 &\tabincell{c}{22, 25, 28, 31, 34,\\ 37, 40, 43, 46, 49}&4096$\times$2160 \\
		\hline
		\hline
		Test \#2 &0& \tabincell{c}{4096$\times$2160, 2880$\times$1620, 2560$\times$1440,\\ 1920$\times$1080 ,1600$\times$900, 1280$\times$720, \\720$\times$480, 320$\times$240}\\
		\hline
		\hline
        \multirow{8}{*}{Test \#3}&\tabincell{c}{22, 25, 28, 31, 34, \\37, 40, 43, 46, 49} &4096$\times$2160 \\ \cline{2-3}
        &\tabincell{c}{22, 25, 28, 31, 34,\\ 37, 40, 43, 46, 49} &3072$\times$1260\\ \cline{2-3}
        &\tabincell{c}{22, 25, 28, 31, 34,\\ 37, 40, 43, 46, 49} &2048$\times$1080\\ \cline{2-3}
        &\tabincell{c}{22, 25, 28, 31, 34,\\ 37, 40, 43, 46, 49} &1024$\times$540\\ \cline{2-3}
		\hline
	\end{tabular}
\end{table}

\subsubsection{Test Protocol} \label{sec:test protocol}
Subjects who wear the HMD, are asked  to stay steady by focusing on a green cross overlaid in the FoV center
 without head and body movement  when performing the tests.
This is to avoid viewing noise when the user randomly shifts their attention in a large area.

As aforementioned, the rendered FoV within the HTC HMD is  110$^{\circ}$-wide horizontally which includes both central and peripheral vision areas. Therefore, we divide each FoV into three main regions, i.e., central vision area (CVA) with one-side eccentricity $\theta$ from 0$^\circ$ to 9$^\circ$, near peripheral area (NPA) with $\theta\in(9^\circ, 30^\circ]$ and the rest from 30$^\circ$ to 55$^\circ$ for the far peripheral area (FPA). We do not discuss the area outside the current FoV in this work.


In general, we combine the double stimulus~\cite{BT500} and just-noticeable-distortion (JND) criteria together. We show consecutive image pairs during the test session. Each pair is displayed for about 5 seconds with a 3-seconds pause to record the subjective JND opinion. There is a 1-minute interval for subjects  to rest their eyes between two different images. For each pair, one is with the uniform quality at its highest quality (e.g., one example is $q$ = $q_{\min}$ and $s$ = $s_{\max}$ ) which is noted as the anchor, and the other one is with the unequal quality settings in central and peripheral areas.  Various quality scales are compared against the anchor to identify the boundary that our HVS could perceive the quality difference. Since we divide the FoV into multi-regions, rating comparison process is performed from the CVA to FPA step-wisely.

Specifically, anchor image is presented at its native spatial resolution $s_{\max}$ with $q = q_{\min}$ (or QP 22).
It is then compared with images that are processed with various $q$s (or equivalent QPs) and $s$s sequentially in a predefined order. Normally, we increase $q$ or reduce $s$ step by step to degrade the image quality, until we finally perceive the quality degradation subjectively - this is referred to as the JND moment. We retrieve the recorded $q$ or $s$ for the image just before the JND moment, i.e., $q$ = $q_c$ or $s$ = $s_c$, as the parameters applied in CVA. In another words, with  $q\leq q_c$ or $s\geq s_c$, we will not sense any perceptual difference  of the image shown in our CVA.


%

Afterwards, we fix the content quality at the CVA of each test sample using the $q_c$ or $s_c$, and degrade the quality  of both the NPA and FPA  together until the subject notices the distortion. We record the corresponding $q$ or $s$, noted as $q_{p_n}$ or $s_{p_n}$, respectively.
Then, we fix the quality of the NPA with the $q_{p_n}$ or $s_{p_n}$, and continue to degrade the quality of the FPA separately till subjects feel the difference perceptually. Similarly, associated $q_{p_f}$ or $s_{p_f}$ are marked.


\subsubsection{Data Post-Processing}

For test \#1, we fix the spatial resolution at $s_{\max}$ and adjust $q$ to obtain the $q_c$, $q_{p_n}$, and $q_{p_f}$ that associate with the JND moment. Similarly,  we derive the $s_c$,  $s_{p_n}$ and $s_{p_f}$ for test \#2.
As discussed earlier, it would contain a large amount of combinations for joint $q$-$s$-impact study. To reduce the complexity, we have constrained a few typical spatial resolutions as shown in Table~\ref{tab:quality_paras}. For each spatial resolution $s$, we follow the similar procedure as test \#1 to record corresponding $q_{s, c}$, $q_{s, p_n}$, and $q_{s, p_f}$.  Such process in test \#3 presumes the separable effects of the quantization and spatial resolution similar as our earlier work~\cite{Xiaokai_TIP,Ma_RQModel}.

It takes about $3\times4\times8\times8/60 = 26$ minutes for each subject to view a test sequence including all $q$-induced quality scales for each immersive image at a particular spatial resolution $s$. Each test sequence is assessed by 40-50 viewers approximately.
We collect all of these raw data (i.e., $q$, $s$) and apply the screening method to remove outliers (so as to reduce the rating noise).
More specifically, we first generate the probability distribution of the $q$ or $s$ at CVA, NPA and FPA respectively, for each image with the data from all subjects, and then calculate the mean $\mu$ standard deviation $\sigma$. For the $j$-th image rated by
the $i$-th subject on $q$-impact, if $| q_{j, i, c} - \mu_{q_{j, c}} | > 2\times\sigma_{q_{j, c}}$, we would exclude this number.
Here,  $\mu_{q_{j, c}}$ and $\sigma_{q_{j, c}}$ are the mean and standard deviation of the corresponding $q$ at central vision area
for $i$-th image measured from all subjects. Furthermore, all the rating data of a subject will be removed if we have found his/her individual rating is excluded for two or more different image samples.  After the data screening, each test sample at a particular quality scale has about 35 valid rating, with the mean of these valid data finally produced as the effective $q$ or $s$ for each immersive image at different visual areas (i.e., CVA, NPA and FPA).



\subsection{Analytical Models}\label{sec:model_detail}

As aforementioned, our HVS exhibits quite different sensation among various areas. Hence, we would like to leverage such characteristics
to implement the unequal quality scales, i.e., reduced quality at peripheral areas, but without perceptual degradation. This would  reduce the image size, given that bit rate consumption is typically decreased when reducing the image quality~\cite{SSIM,Ma_RQModel}.
Therefore, quantitative models that expresses the perceptual quality considering the peripheral vision impact are highly desired.
This section details the model development. 
More specifically, we first explore how central and peripheral vision influence the distinguishability of image's spatial resolution $s$or quantization stepsize $q$ separately, and then extend the study on joint impacts of the quantization and spatial resolution.
Together with our previous work~\cite{VR_spatial_IC3D,Xiaokai_TIP,pv_mobileQSTAR}, we finally provide a closed-form
perceptual quality model to explicitly explain the image quality (in terms of the MOS) with the consideration of unequal impact in the central and peripheral vision areas.

\begin{figure}[t]
	\centering
	\subfigure[Attic]{\includegraphics[scale=0.5]{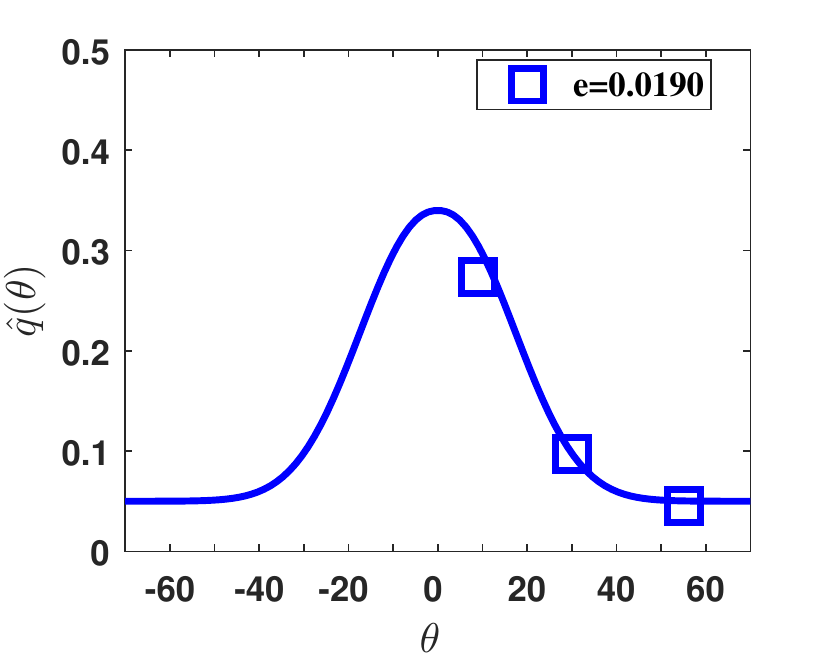}}
	\subfigure[Temple]{\includegraphics[scale=0.5]{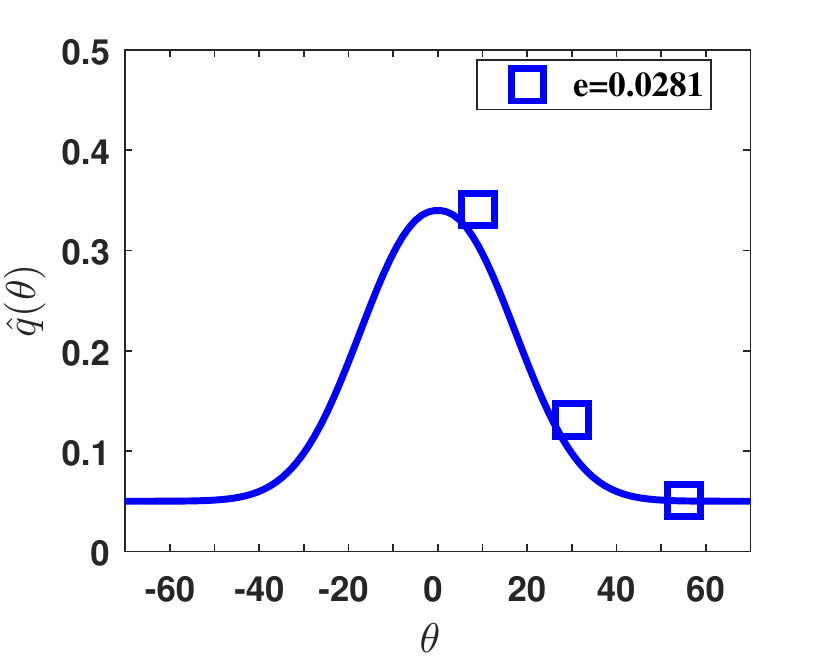}}
	\subfigure[Ship]{\includegraphics[scale=0.5]{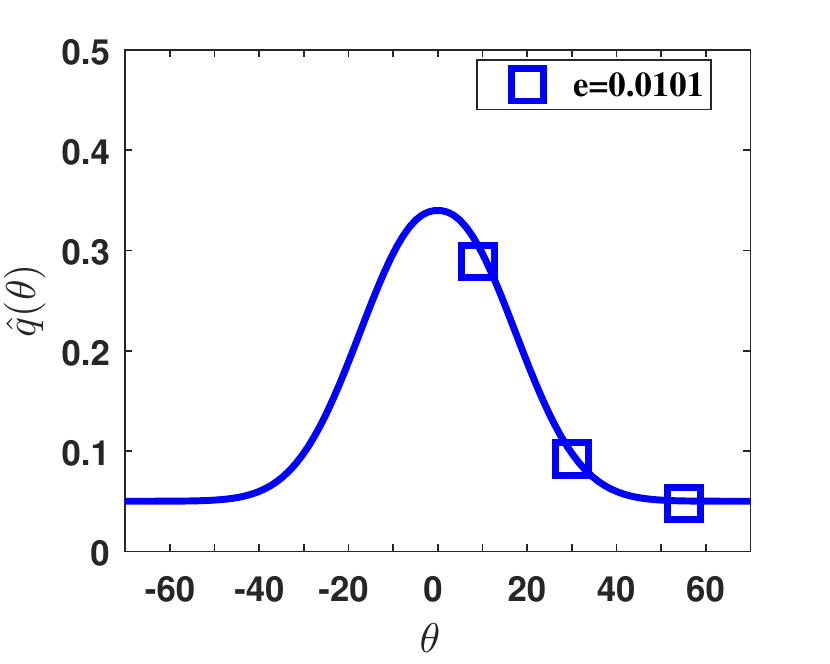}}
	\subfigure[Train]{\includegraphics[scale=0.5]{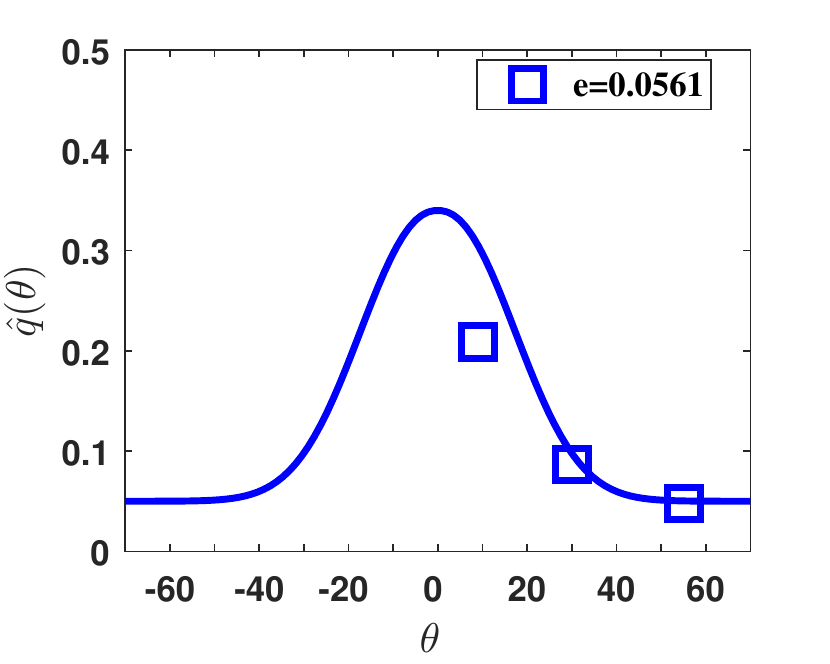}}
	\subfigure[Beach]{\includegraphics[scale=0.5]{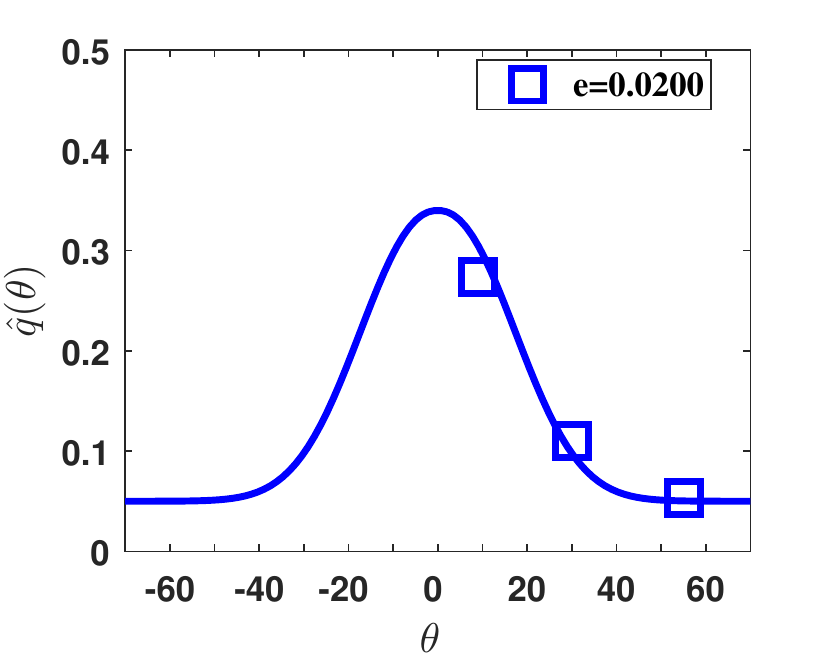}}
	\subfigure[Sculpture]{\includegraphics[scale=0.5]{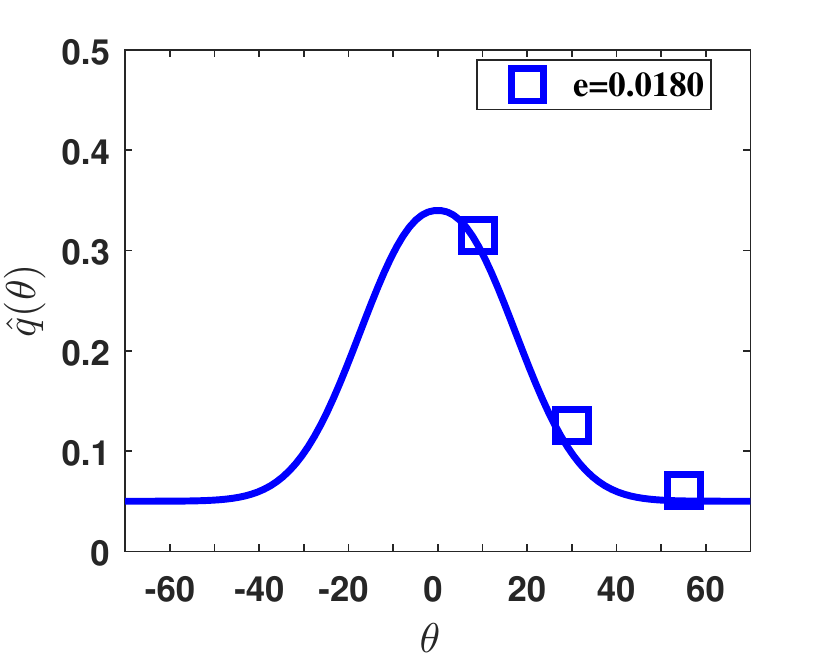}}
	\subfigure[Football]{\includegraphics[scale=0.5]{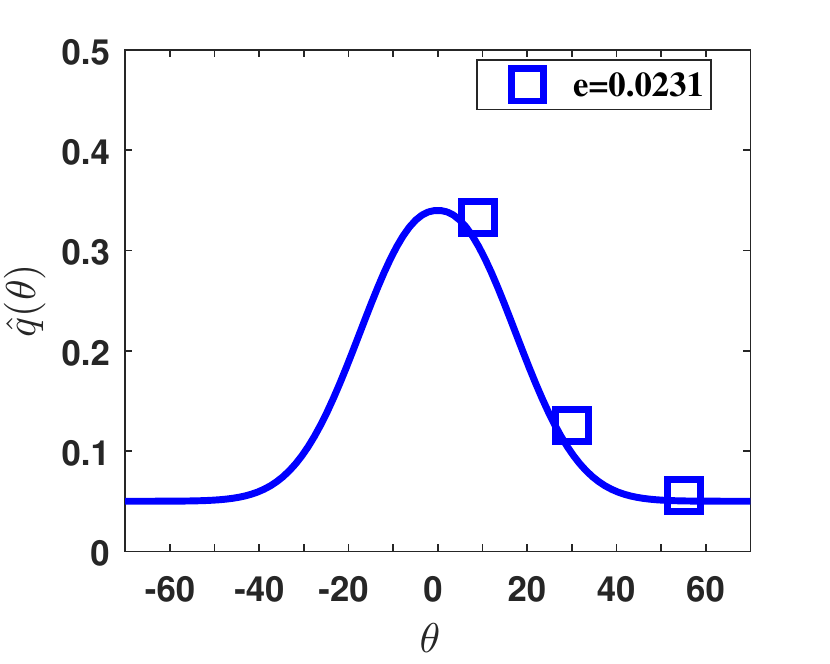}}
	\subfigure[Desert]{\includegraphics[scale=0.5]{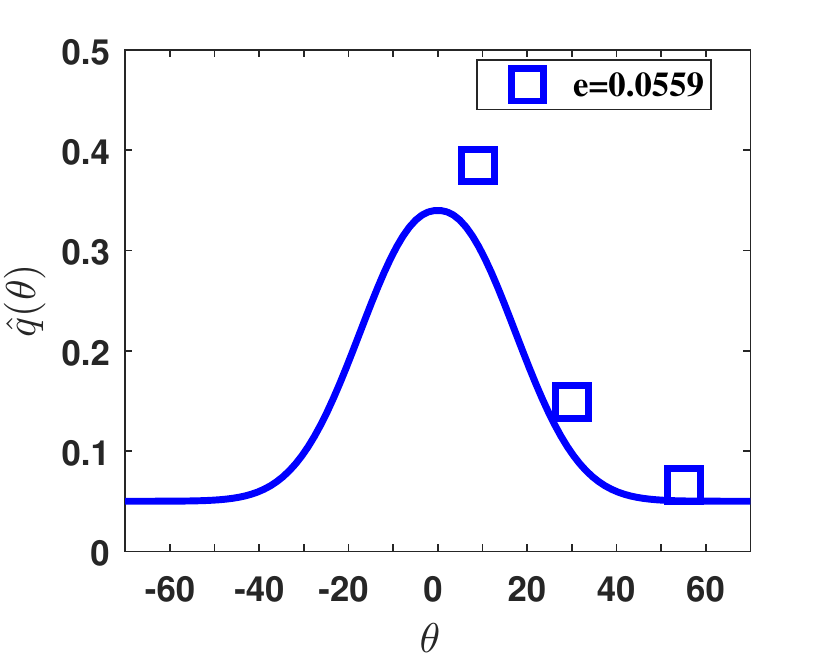}}
	\caption{Normalized quantization versus eccentric angle $\theta$. $e$ represents the root mean square error (RMSE). Parameters
	are fixed for all image content. Discrete points are measured data, while continuous curve is fitted model.}
	\label{fig:normalized_data_q}
\end{figure}

\begin{figure}[t]
	\centering
	\subfigure[Attic]{\includegraphics[scale=0.5]{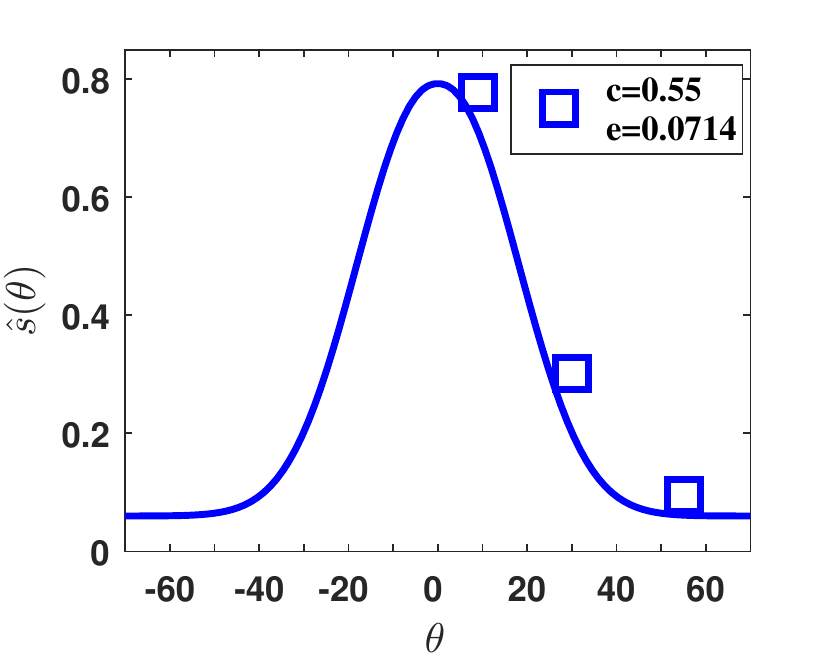}}
	\subfigure[Temple]{\includegraphics[scale=0.5]{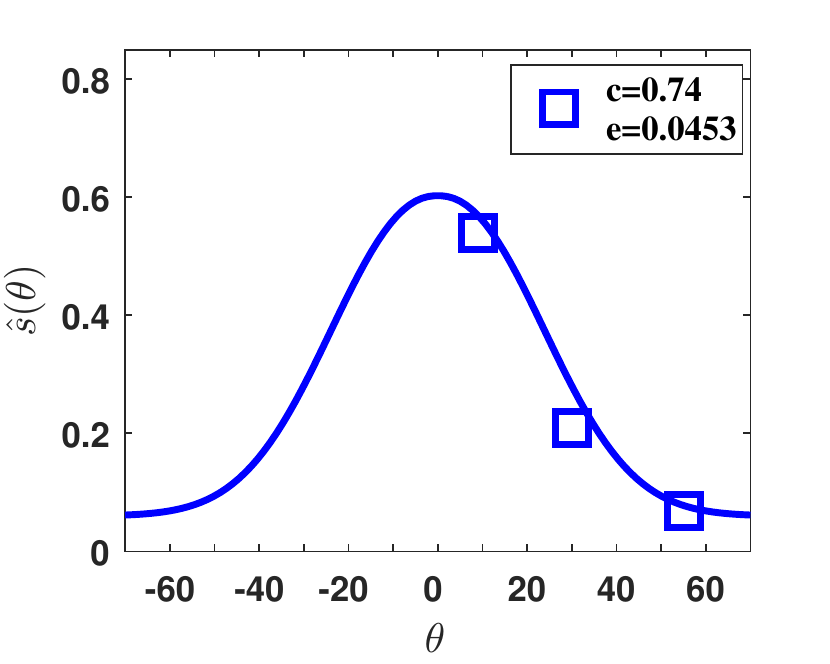}}
	\subfigure[Ship]{\includegraphics[scale=0.5]{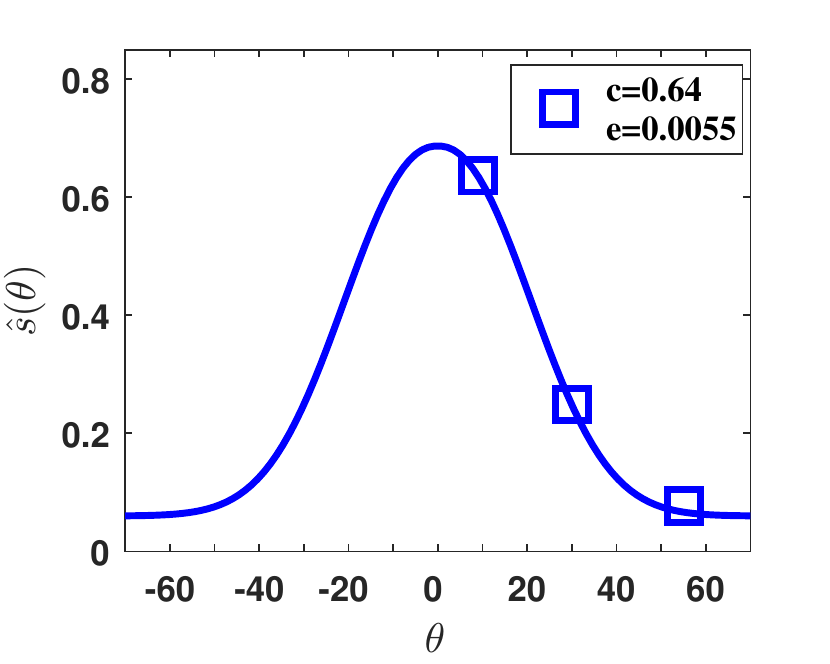}}
	\subfigure[Train]{\includegraphics[scale=0.5]{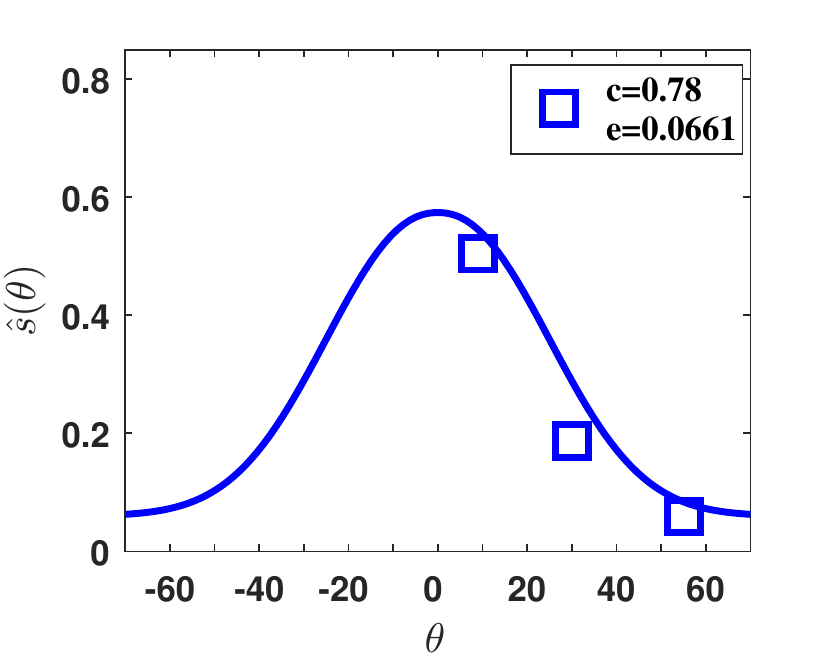}}
	\subfigure[Beach]{\includegraphics[scale=0.5]{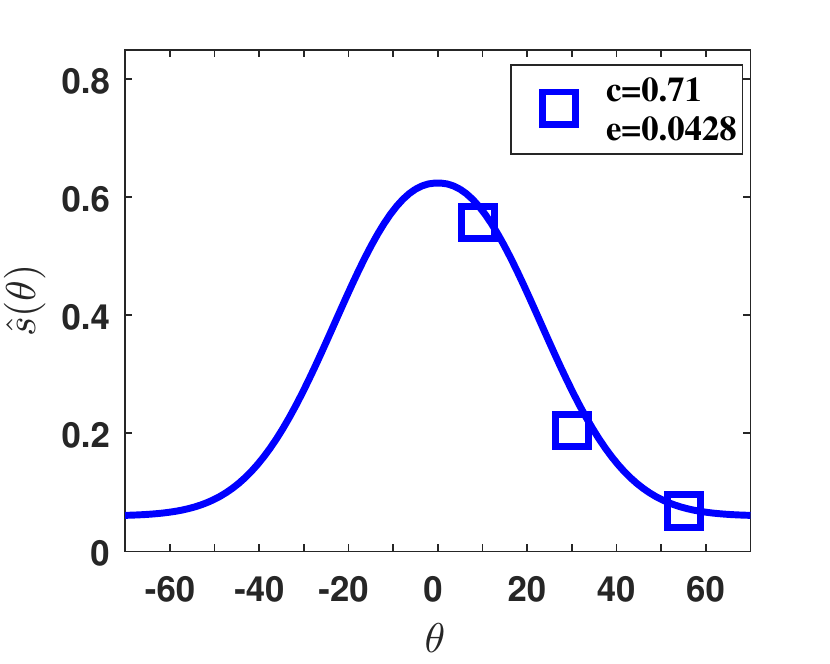}}
	\subfigure[Sculpture]{\includegraphics[scale=0.5]{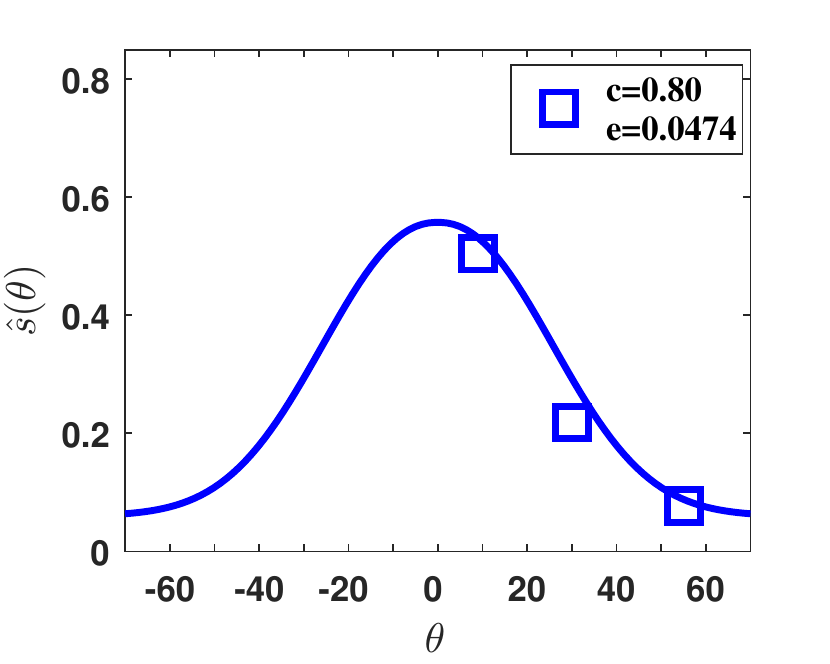}}
	\subfigure[Football]{\includegraphics[scale=0.5]{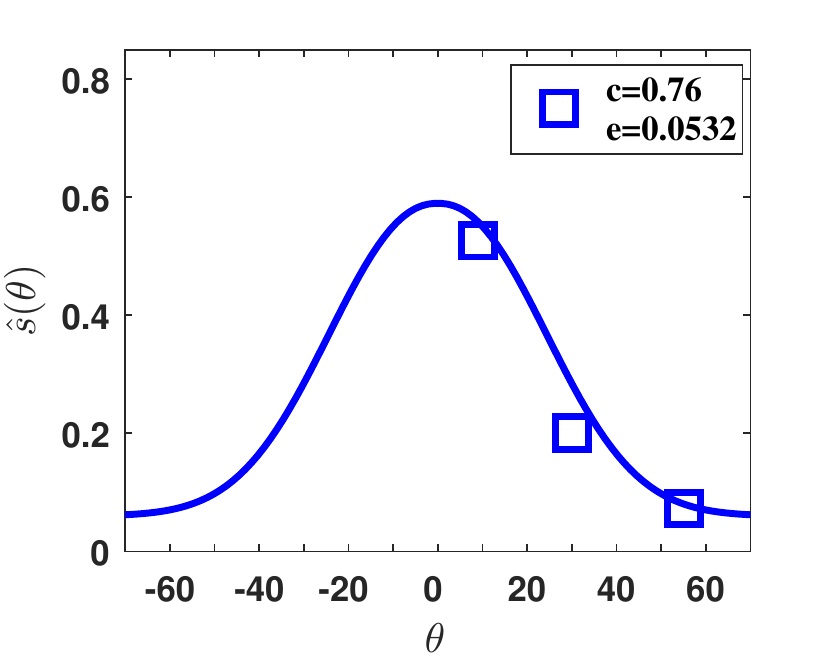}}
	\subfigure[Desert]{\includegraphics[scale=0.5]{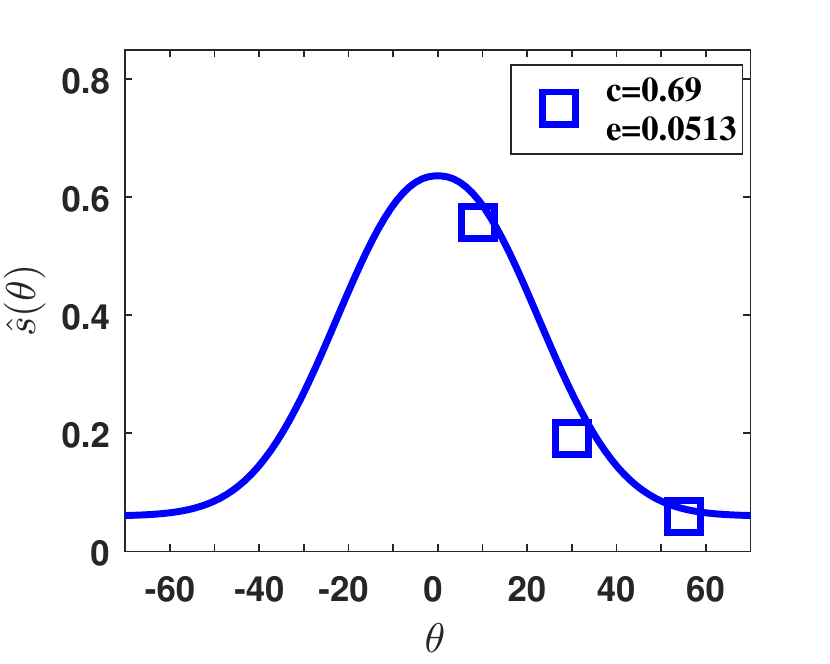}}
	\caption{Normalized spatial resolution versus eccentric angle $\theta$. $e$ represents the root mean square error (RMSE).
	Parameters except $c$ are fixed, while $c$ is content dependent. Discrete points are measured data, while continuous curve is fitted model.}
	\label{fig:normalized_data_s}
\end{figure}

\subsubsection{Separate Impact of Quantization and Spatial Resolution} \label{sssec:separable_q_s_impact}
To simplify the model development,
 $q$ and $s$ are first normalized using $q_{\min}$ = 8 (or QP 22) and $s_{\max}$ = $4096\times2160$, i.e.,
$\hat{q} = q_{\min}/q$, and $\hat{s}= s/s_{\max}$, respectively.
When $q$ = 64, the corresponding $\hat{q}$ is 0.125; while $s$ = $2048\times1080$, $\hat{s} = 0.25 $.
Figure~\ref{fig:normalized_data_q} and~\ref{fig:normalized_data_s} show the collected $\hat{q}$ and $\hat{s}$ with
respect to the eccentric angle $\theta$.

We have found that an unified parametric generalized Gaussian function could explain the $\hat{q}$ and $\hat{s}$ very well, i.e.,
\begin{equation}
	\frac{1}{{c\sqrt {2\pi } }} \times {e^{ - \frac{{\left| {{{\left( {b\cdot\theta} \right)}^a }} \right|}}{{2{c^2}}}}} + d,
	\label{eq:q_s_model}
\end{equation} where $a$, $b$, $c$, $d$ are control parameters derived using the least-squared criteria by fitting the measured points and hypothesized analytical model in Eq.~\eqref{eq:q_s_model}. This model~\eqref{eq:q_s_model} also coincides with the density distribution of photoreceptors $\rho(\theta)$, where Laplacian distributed $\rho(\theta)$ is a special case of the generalized Gaussian function.

Fitted parameters for $\hat{q}(\theta)$ and $\hat{s}(\theta)$ are shown in Table~\ref{tab:paras}, respectively.  Note that parameters are different for $q$ and $s$ due to the reason that adapting the $q$ infers the high frequency information loss of compression while varying the $s$ implies the loss of some low frequency content.
\begin{table}[b]
	\centering
	\caption{Parameters in peripheral vision model for $\hat{q}(\theta)$ or $\hat{s}(\theta)$}
	\label{tab:paras}
	\begin{tabular}{|c|c|c|c|c|}
		\hline
		& $a$ & $b$ & $c$ & $d$\\
		\hline
		$\hat{q}(\theta)$ &2.2&0.08&1.38&0.05\\
		\hline
		$\hat{s}(\theta)$ &2.2&0.033&$c(\bf x)$&0.06\\
		\hline
	\end{tabular}\\
	$\bf x$ in $c(\bf x)$ are content features extracted from the image.
\end{table}

Parameter $a$ is identical as it reflects the decay speed of the quality perception with respect to the increase of the $\theta$. This is mainly because it is determined by the density of the cones of human retina.
$b$ is correlated with the quality degradation factors. As aforementioned, $q$ would introduce compress-induced high frequency information loss while $s$ brings the loss of some low frequency content due to the spatial re-sampling.

Parameter $c$ is generally content-dependent.  But for $\hat{q}$, we can still use the fixed $c$ for all images. It is suspected that current Vive display does not offer sufficient pixel density (e.g., pixel per inch or PPI) to accurately reflect the compression-induced amplitude variations per pixel.  But for spatial re-sampling,  pixels are restored with predefined filers, which significantly differ from the original pixel in native 4096$\times$2160 resolution.  

Parameter $d$ indicates the model limits based on our intuition. For example, we could not perceive any difference when $\theta$ goes to infinity (i.e., the number of the photoreceptor goes to zero).  For $q$,  we could apply the $q = q_{\max} = 228$ (with corresponding QP 51 used in H.264/AVC and HEVC~\cite{AVC,HEVC}), implying $d$ = 0.03. But the worst image quality that subjects can distinguish is with  QP 48, it's  suitable to set $d$= 0.05; For $s$, it is impractical to have $s$ = 0 for rendering, thus we set $d$ = 0.06 with the least model prediction error.

Furthermore, we introduce how to predict parameter $c$ for model $\hat{s}(\theta)$.  Intuitively, image quality is mainly determined by its  spatial complexity, color distribution, and local orientation.  Through careful examination, we have found that $c$ could be predicted by the linear combination of the $\rho_{c_{\tt SI}}$, $\rho_{\mu_{I}}$ and $\rho_{\mu_{\gamma_v}}$, i.e.,
\begin{equation}
c =  - 0.002\cdot\rho_{c_{\tt SI}} + 0.4342\cdot\rho_{\mu_{I}}+ 3.9029\cdot\rho_{\mu_{\gamma_v}} + 0.2557,
	\label{parameter-c-prediction}
\end{equation}
 where $\rho_{c_{\tt SI}}$ is the SI of the central vision area of an image. (This is because we constrain the saliency region in the central vision.)  $\rho_{\mu_{I}}$ is the averaged intensity of the image in HSI color space~\cite{hsi_color_space}.  $\rho_{\mu_{\gamma_v}}$ refers to the intensity of the vertical orientation which is calculated using a 3$\times$3 Gabor filter~\cite{wiki:gabor}.

\subsubsection{Joint Impacts of Quantization and Spatial Resolution}
This section investigates the joint impacts of the quantization and spatial resolution on the perceptual quality with respect to the eccentricity $\theta$. Generally speaking, the joint impacts are not indeed separable. However, motivated by our previous work~\cite{Ma_RQModel, pv_mobileQSTAR, R-STAR}, we still attempt to enforce the separable effects for $q$ and $s$. Towards this purpose, we have performed the test \#3 shown in Table~\ref{tab:quality_paras}, where the $q$-impact is studied at different spatial resolutions. To reduce the overall rating duration, we use a few typical spatial resolutions, but still allow ten distinct $q$s to cover a variety of quality scales.

\begin{figure*}[t]
	\centering
	\subfigure[Attic]{\includegraphics[scale=0.5]{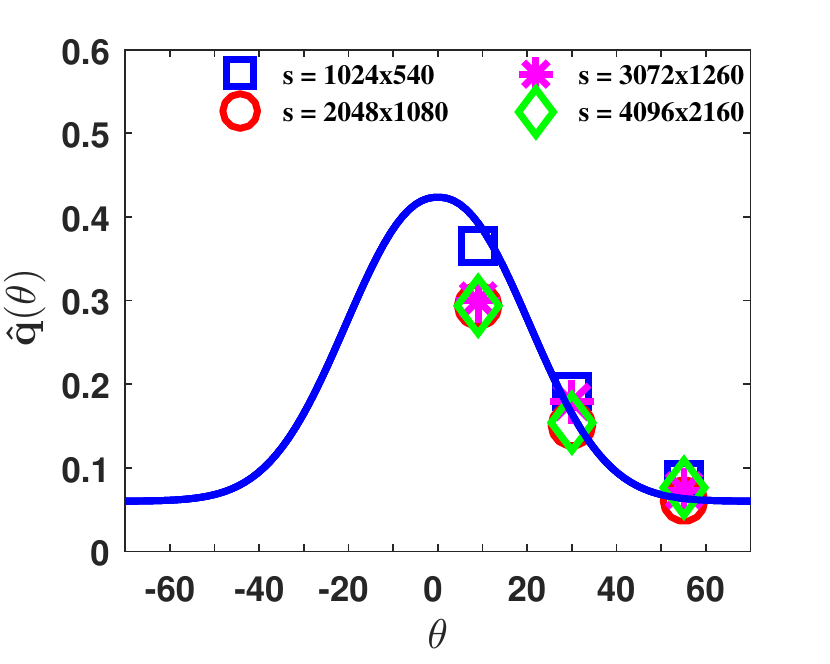}}
	\subfigure[Temple]{\includegraphics[scale=0.5]{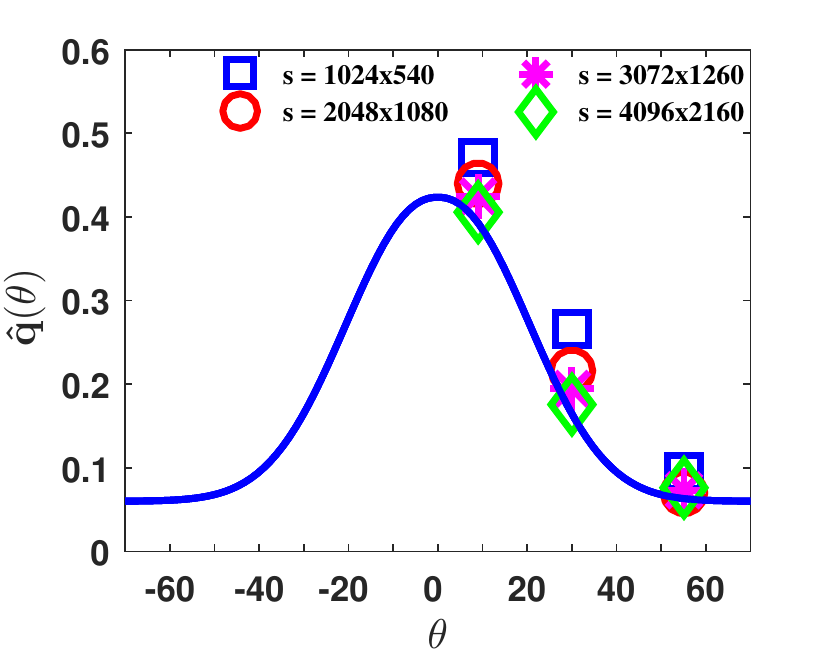}}
	\subfigure[Ship]{\includegraphics[scale=0.5]{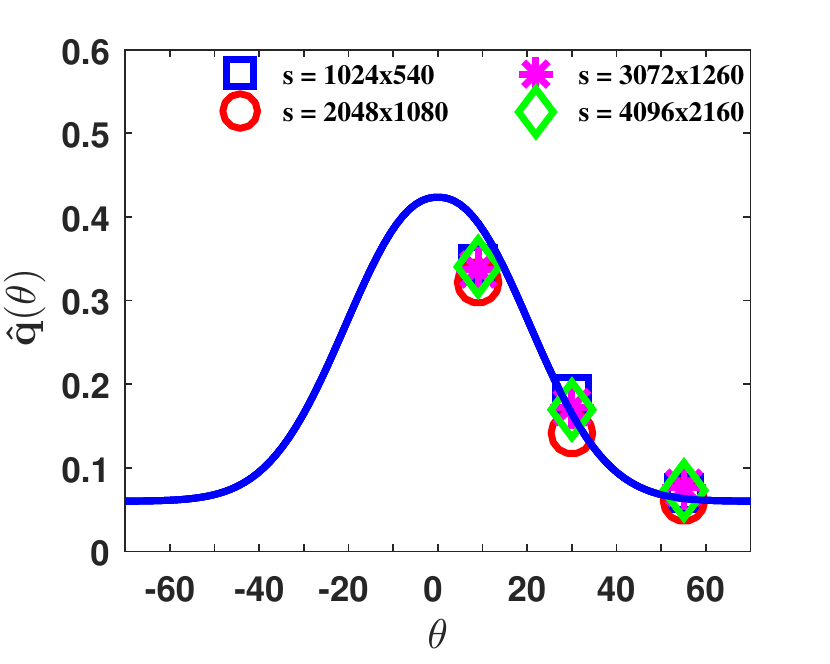}}
	\subfigure[Train]{\includegraphics[scale=0.5]{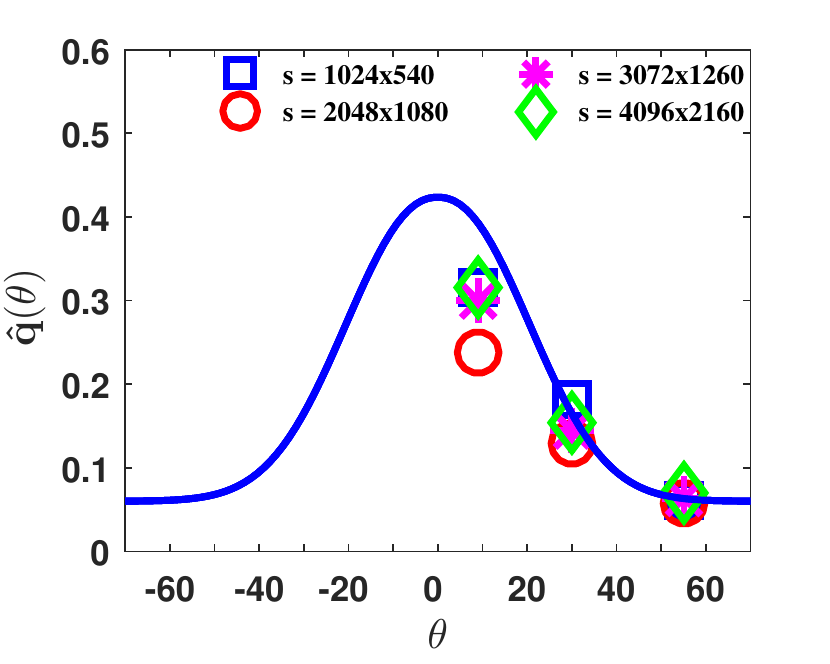}}
	\subfigure[Beach]{\includegraphics[scale=0.5]{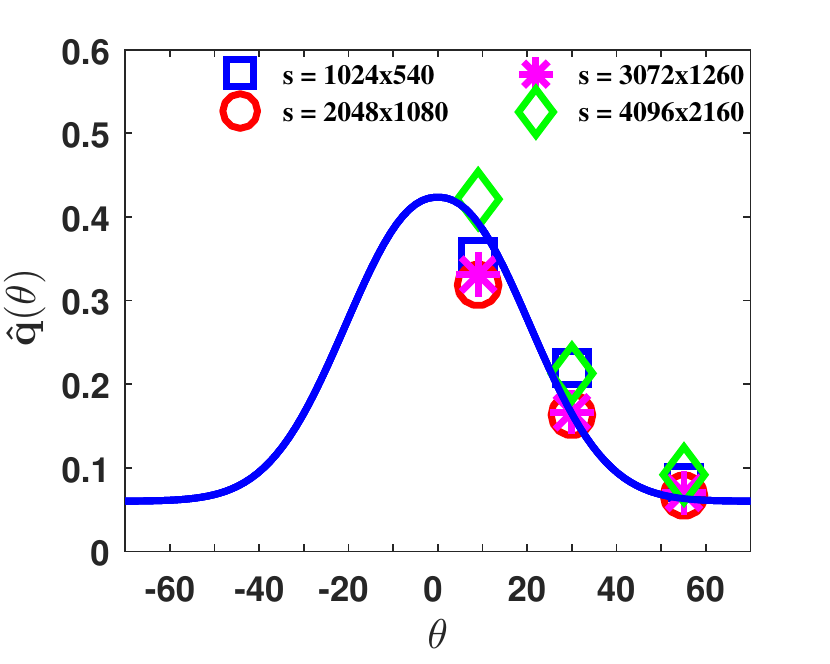}}
	\subfigure[Sculpture]{\includegraphics[scale=0.5]{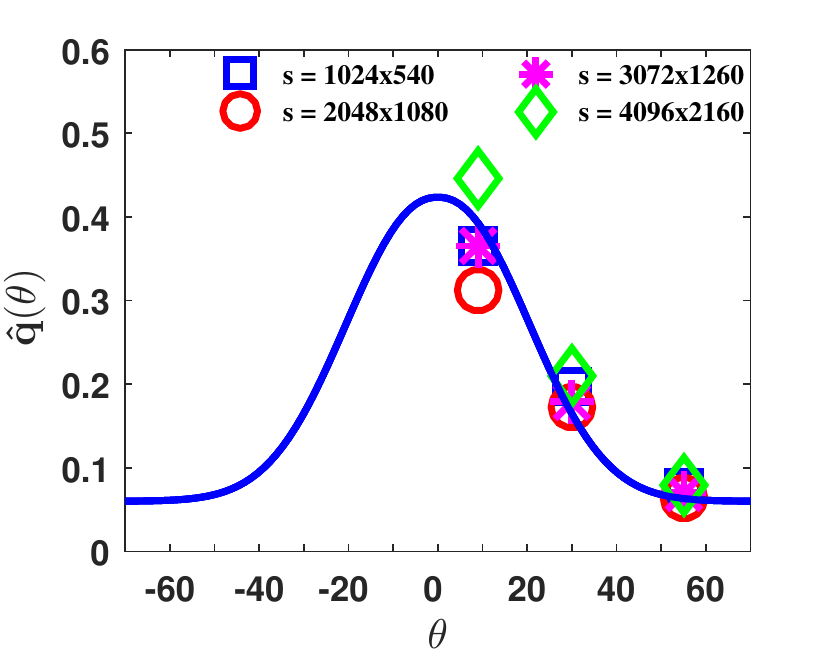}}
	\subfigure[Football]{\includegraphics[scale=0.5]{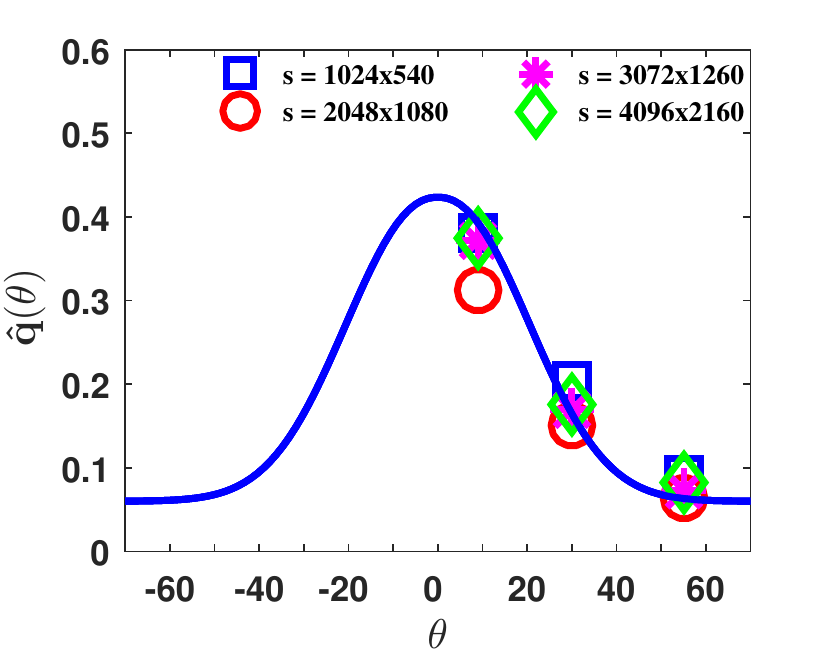}}
	\subfigure[Desert]{\includegraphics[scale=0.5]{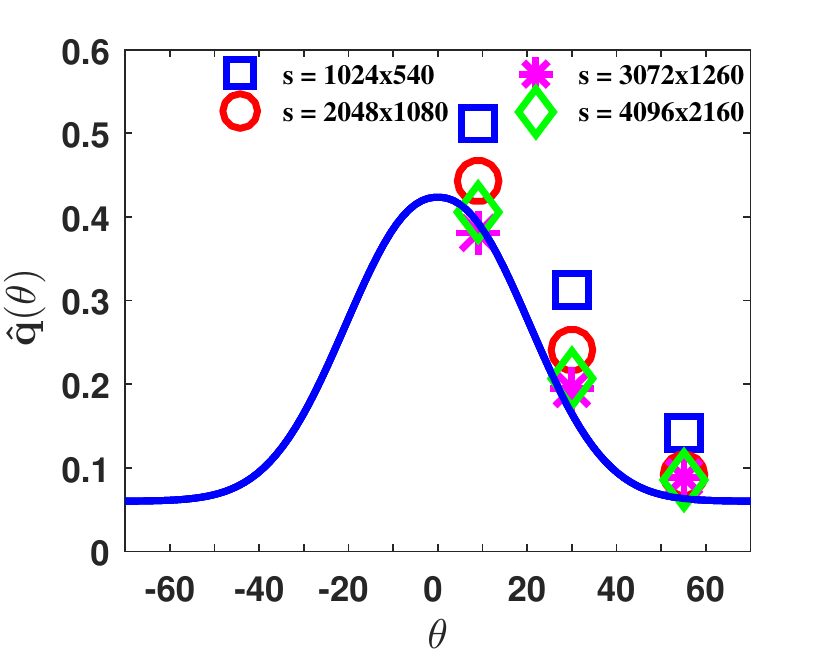}}
	\caption{Normalized $\hat{q}(\theta)$ at different spatial resolutions. Discrete points are measured data; while the curve is fitted model.}
	\label{fig:qs_plot}
\end{figure*}

 We plot the normalized $\hat{q}(\theta)$ at different spatial resolution $s$ in Fig.~\ref{fig:qs_plot}. It is found that discrete measurements are almost overlapped for different spatial resolutions. This implies that a single analytical model may be sufficient to explain the $q$-impact at different $s$. Nevertheless, we directly fit the discrete  $\hat{q}(\theta)$s using Eq.~\eqref{eq:q_s_model}, first assuming the independent parameters at different spatial resolution, and then enforcing the same parameters for all spatial resolutions, via the least squared error criteria. Parameters are listed in Table~\ref{tab:multi_s_q}. As seen, we could definitely pursue the least error with $s$ dependent parameters, but a model with fixed parameters is easier for future application deployment. On the other hand, the error is within reasonable range for fixed parameters setup in comparison to the $s$ adaptive parameters. Thus, we propose to apply the fixed parameters for the following discussion. As will be unfolded
 in cross-validation, even with the fixed parameters, our model still offers quite impressive performance for subjective quality estimation together
 with our preliminary studies in~\cite{Xiaokai_TIP}.


%


 \begin{table}[b]
    \renewcommand\arraystretch{1.25}
	\centering
	\caption{Parameters  for $\hat{q}(\theta)$ at different spatial resolution.  Fitting error is represented using RMSE.}
	\label{tab:multi_s_q}
	\begin{tabular}{|c|c|c|c|c|c|c|}
		\hline
		& $s$ & $a$ & $b$ & $c$ & $d$ & $e$ \\
		\hline
        \multirow{4}{*}{$\hat{q}(s, \theta)$}& 4096$\times$2160 & 2.2 & 0.05 & 1.2 & 0.05 & 0.0330\\ \cline{2-7}
        & 3072$\times$1260 & 2.2 & 0.05 & 1.3 & 0.05 & 0.02365 \\ \cline{2-7}
        & 2048$\times$1080 & 2.4 & 0.06 & 1.2 & 0.06 &0.04789 \\ \cline{2-7}
        & 1024$\times$540 & 2.4 & 0.05 & 1.1 & 0.08 & 0.04733\\
		\hline
        $\hat{q}(\theta)$& & 2.2 & 0.055 &1.1  & 0.06 & 0.04567   \\
        \hline
	\end{tabular}
\end{table}

\subsubsection{The Overall Quality Model considering the Joint Impacts of Quantization and Spatial Resolution }
We could easily derive the appropriate quality control factors $q$ and $s$ with respect to the $\theta$ following the above discussion.
However, in practice,  we are often required to provide the subjective quality quantitatively,
i.e., relating the quality control factors to the perceptual quality (typically represented using the MOS).
Fortunately, our previous attempts in~\cite{Ma_RQModel, spatial_temporal, pv_mobileQSTAR} have
investigated the impacts of the spatial resolution, and quantization on conventional image/video
that are rendered with a very limited FoV. Recently, we have extended this study to the immersive image
displayed using the HMD with a 110$^\circ$~\cite{VR_spatial_IC3D,Xiaokai_TIP} FoV.  All of these studies have been
conducted for the image/video with the uniform quality, even for the immersive image.
Moreover, these works
suggest that the impacts of the spatial resolution and quantization on the perceptual quality are generally separable for practical application.

Combining the models in~\cite{Xiaokai_TIP} and Eq.~\eqref{eq:q_s_model}, we could finally reach a closed-form function at
  \begin{align}
   Q(\hat{s},\hat{q}) = {Q_{\max }} \cdot {\frac{{1 - {e^{ - \alpha\cdot{\hat{s}^{^{0.7}}}}}}}{{1 - {e^{ - \alpha }}}}} \cdot \frac{{1 - {e^{ - \beta (s) \cdot\hat{q}}}}}{{1 - {e^{ - \beta (s)}}}},
  \end{align} where $Q_{\max}$ is the averaged MOS of the image at $s_{\max}$ and $q_{\min}$, and parameter $\alpha$ and $\beta(s)$ can be predicted via  features of the image content itself. As demonstrated in~\cite{Xiaokai_TIP},
  $Q_{\max}$ can be set as a constant, i.e., 86.  This also fits our intuition that users might give  similar MOS for the same high fidelity content,
  as long as they have corrected visual sensation.

  For the scenario that only spatial resolution adapts with fixed quantization, $Q(\hat{s}, \hat{q})$ is deducted into $Q(\hat{s})$.
 Following the model in~\eqref{eq:q_s_model}, we could assign the different spatial resolutions
 in central and peripheral vision areas, but still offering the same perceptual quality as the one using the uniform spatial resolution everywhere.
 Intuitively, we use the highest spatial resolution at the central vision area, but reduced spatial resolution in the periphery.
 This implies that the perceptual quality of the immersive image with non-uniform spatial resolution in the central and peripheral vision areas
 is determinted by the quality in the central vision area, i.e.,
 \begin{align}
 Q(\hat{s}(\theta)) = Q(\hat{s}(\theta_c)) =  {Q_{\max }} \cdot {\frac{{1 - {e^{ - \alpha\cdot{\hat{s}_c^{^{0.7}}}}}}}{{1 - {e^{ - \alpha }}}}}, \label{eq:s_model}
 \end{align} with $s_c = \hat{s}(\theta_c)$ representing the spatial resolution in the central vision area within the eccentric $\theta_c$.

 Similarly, we could reach at
  \begin{align}
 Q(\hat{q}(\theta)) = Q(\hat{q}(\theta_c)) =  {Q_{\max }}\cdot\frac{{1 - {e^{ - \beta (s_{\max}) \cdot\hat{q}_c}}}}{{1 - {e^{ - \beta (s_{\max})}}}}, \label{eq:q_model}
 \end{align} with $q_c = \hat{q}(\theta_c)$ representing the quantization stepsize in the central vision area within the eccentric $\theta_c$,
 for the case that only quantization stepsize $q$ varies at the native spatial resolution $s_{\max}$.

Furthermore,  as revealed in aforementioned study on the joint impacts of the quantization and spatial resolation, the $q$-impact can be
fixed for all spatial resolutions. Thus, we would derive the perceptual quality of the image  as
  \begin{align}
   Q(\hat{s},\hat{q}(\theta)) &= Q(\hat{s},\hat{q}(\theta_c)) \nonumber\\
   &= {Q_{\max }} \cdot {\frac{{1 - {e^{ - \alpha\cdot{\hat{s}^{^{0.7}}}}}}}{{1 - {e^{ - \alpha }}}}} \cdot \frac{{1 - {e^{ - \beta (s) \cdot\hat{q}_c}}}}{{1 - {e^{ - \beta (s)}}}}, \label{eq:q_s_MOSmodel}
\end{align} where $s$ is the current spatial resolution and $q_c$ is the quantization stepsize in the central vision area.

\begin{figure*}[t]
\centering
\subfigure[Studio]{\includegraphics[scale=0.073]{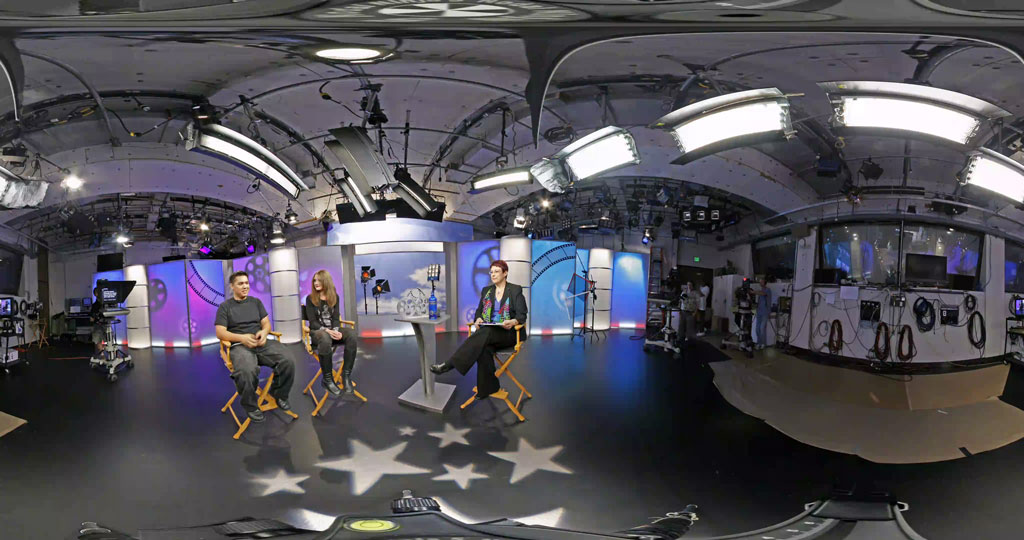}}
\subfigure[Man]{\includegraphics[scale=0.073]{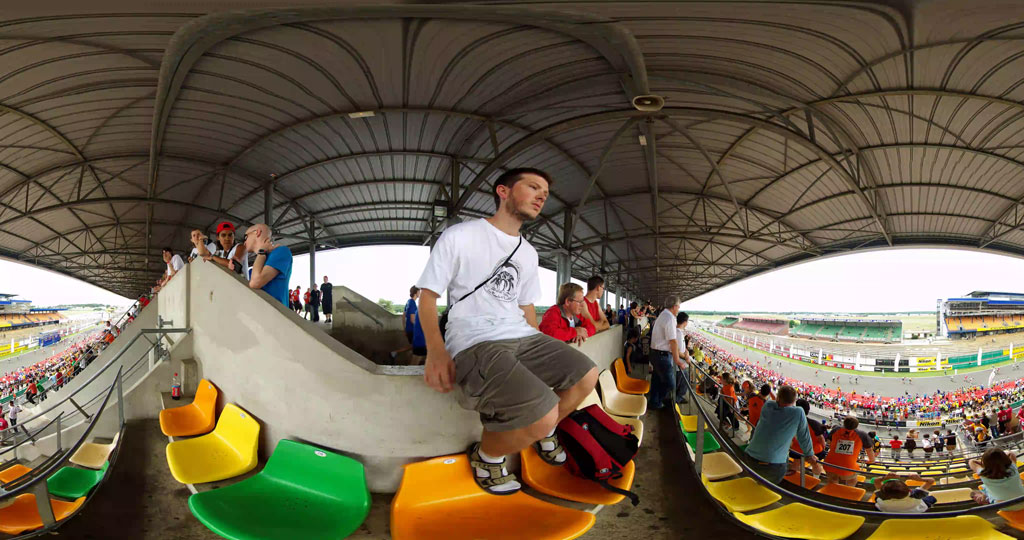}}
\subfigure[Diving]{\includegraphics[scale=0.073]{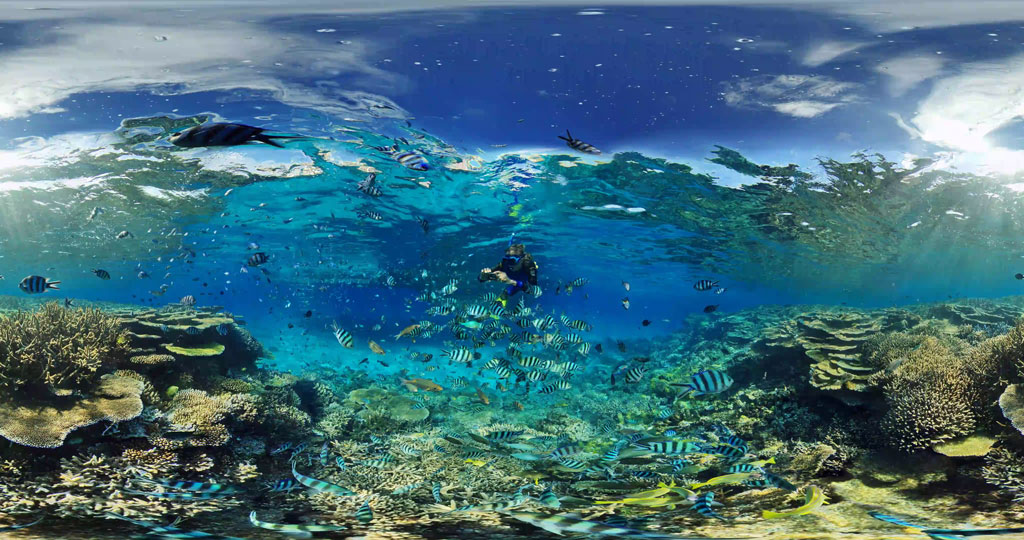}}
\subfigure[Singer]{\includegraphics[scale=0.073]{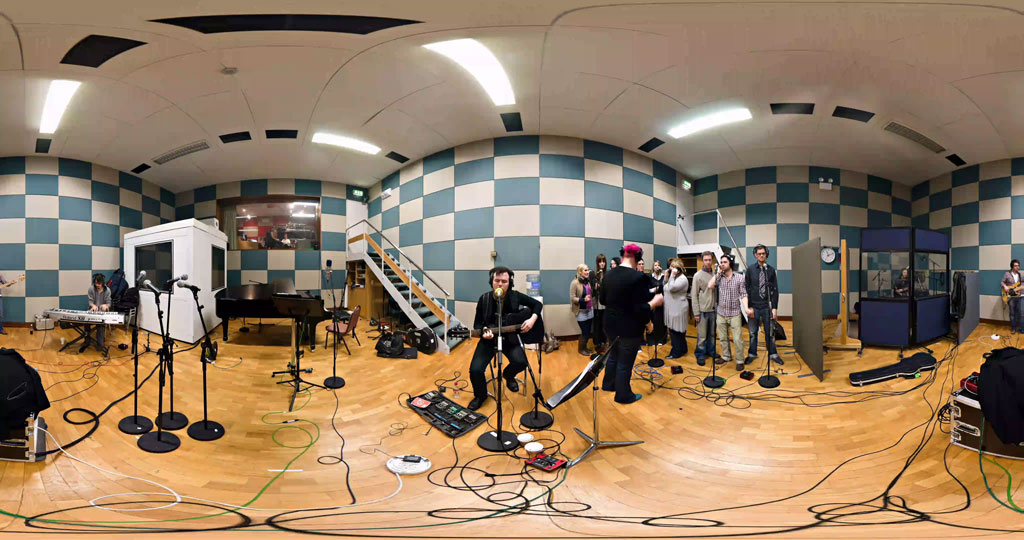}}
\subfigure[Gym]{\includegraphics[scale=0.073]{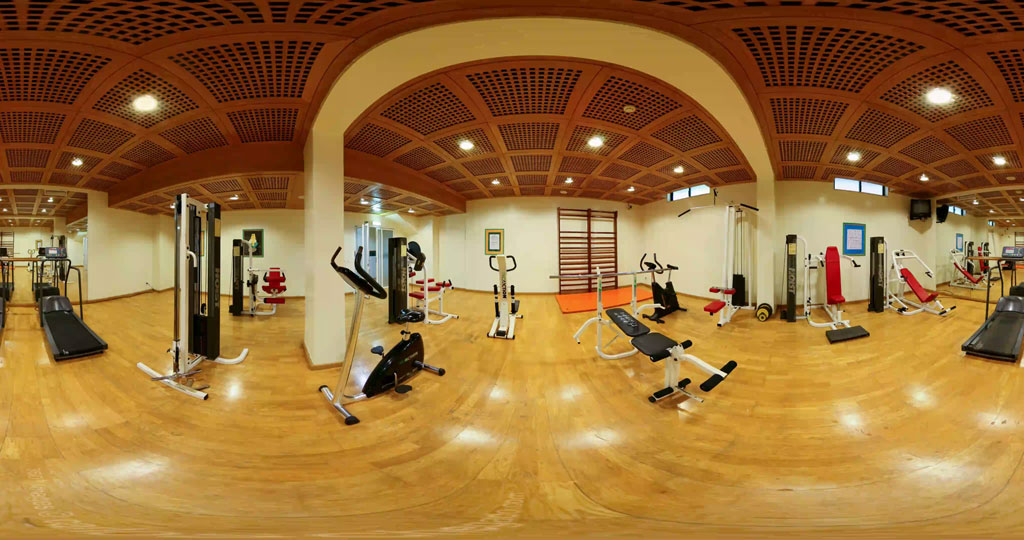}}
\subfigure[Fish]{\includegraphics[scale=0.073]{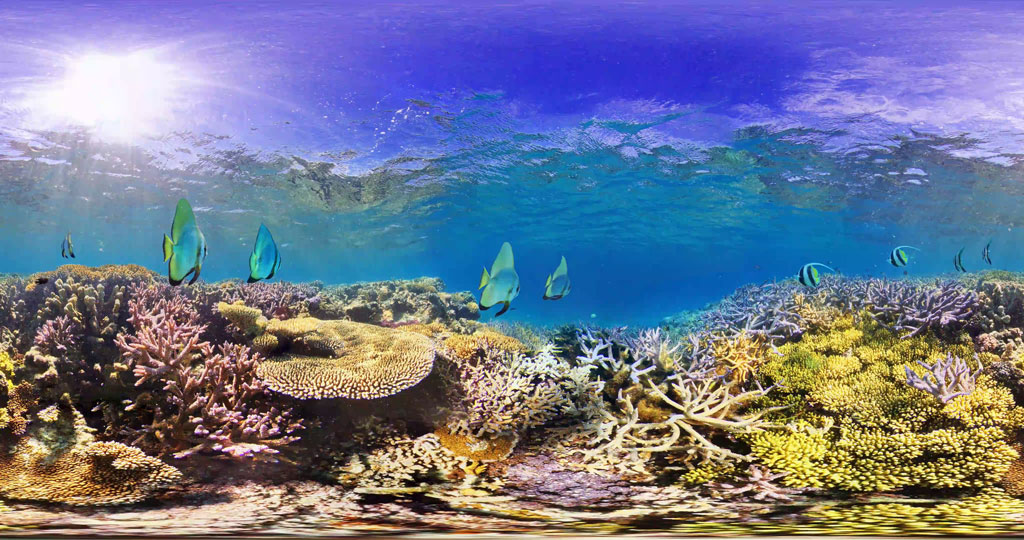}}\\
\subfigure[Arena]{\includegraphics[scale=0.073]{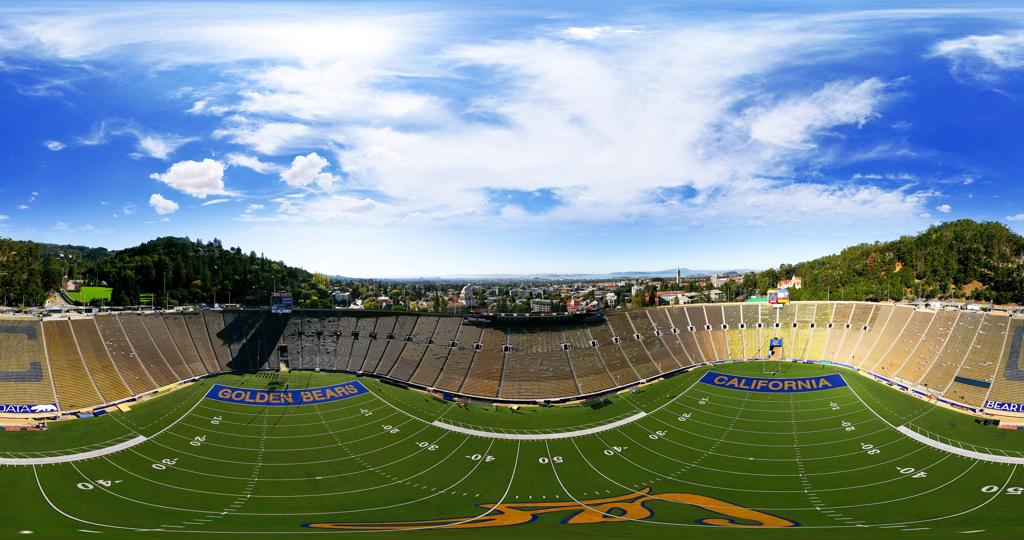}}
\subfigure[Garden]{\includegraphics[scale=0.073]{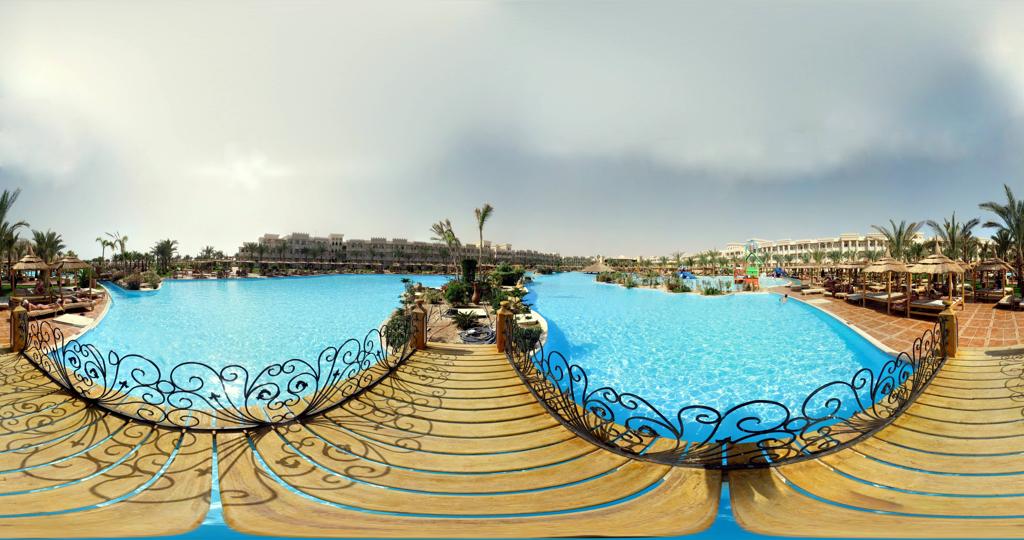}}
\subfigure[Conference]{\includegraphics[scale=0.073]{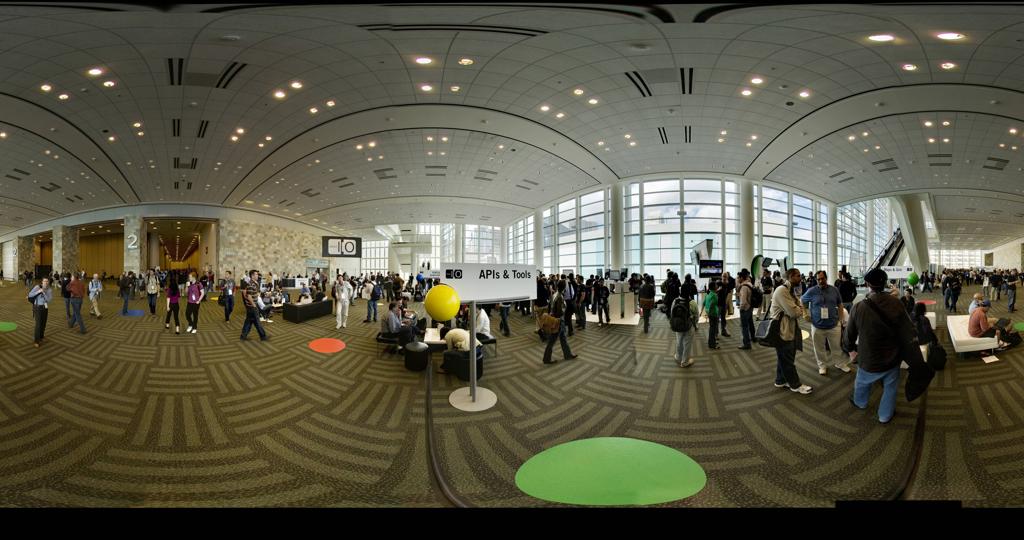}}
\subfigure[Church]{\includegraphics[scale=0.073]{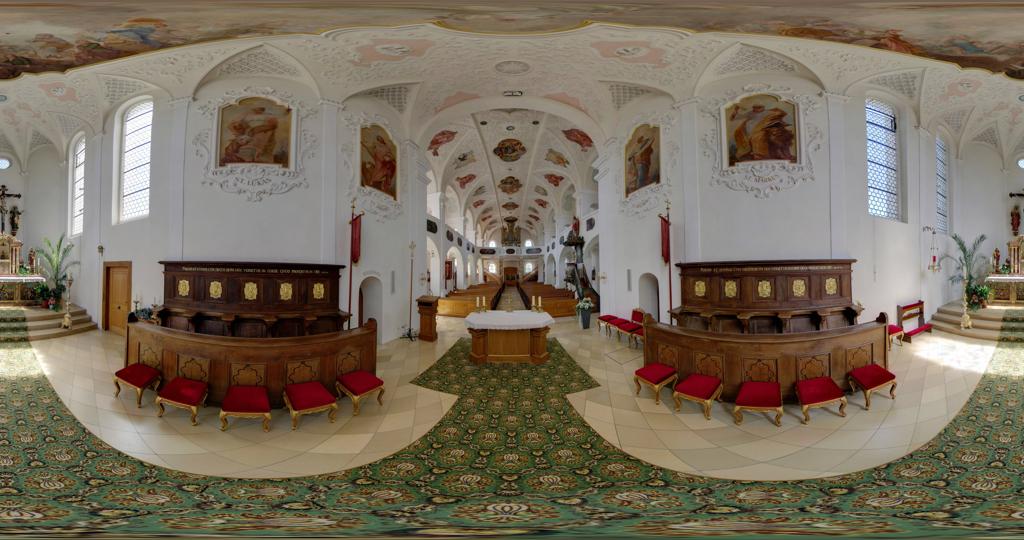}}
\subfigure[Wedding]{\includegraphics[scale=0.073]{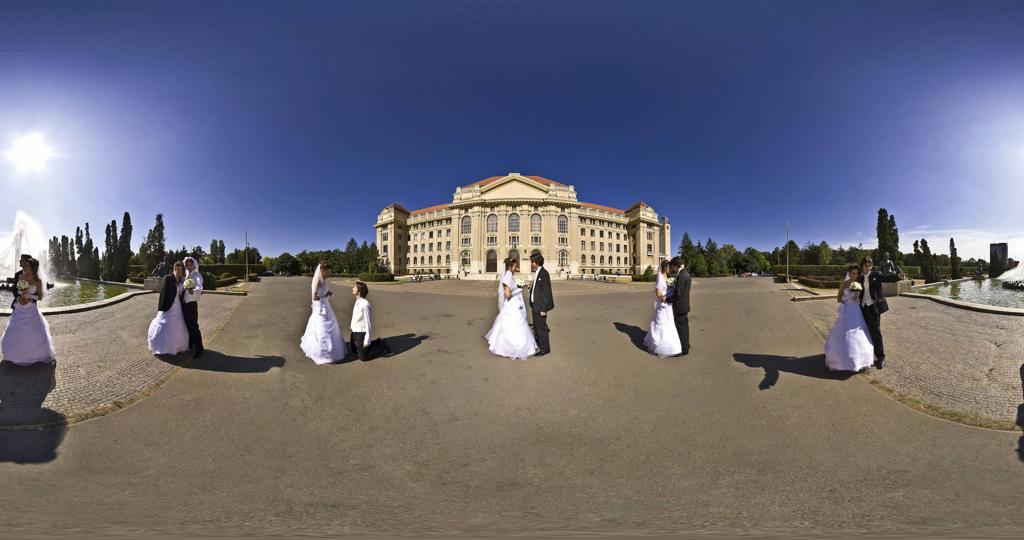}}
\subfigure[House]{\includegraphics[scale=0.073]{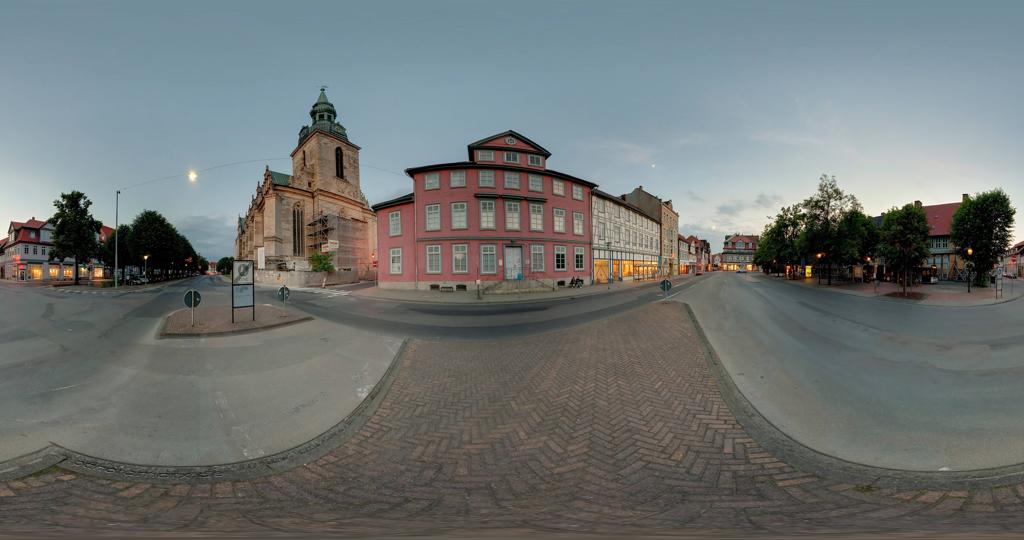}}
\caption{Immersive images used for independent cross-validation that are randomly selected from SUN360 database~\cite{xiao2012recognizing} to cover a variety of  scenarios.}
\label{fig:sourcedata}
\end{figure*}

\section{Independent Model Cross-Validation } \label{sec:cross-validation}
This section details the cross-validation for our proposed analytical models.

\subsection{Cross-Validation of Separate Impacts of Quantization and Spatial Resolution} \label{ssec:x_valid_q_s_impact}
\begin{figure}[b]
	\centering
	\includegraphics[scale=0.17]{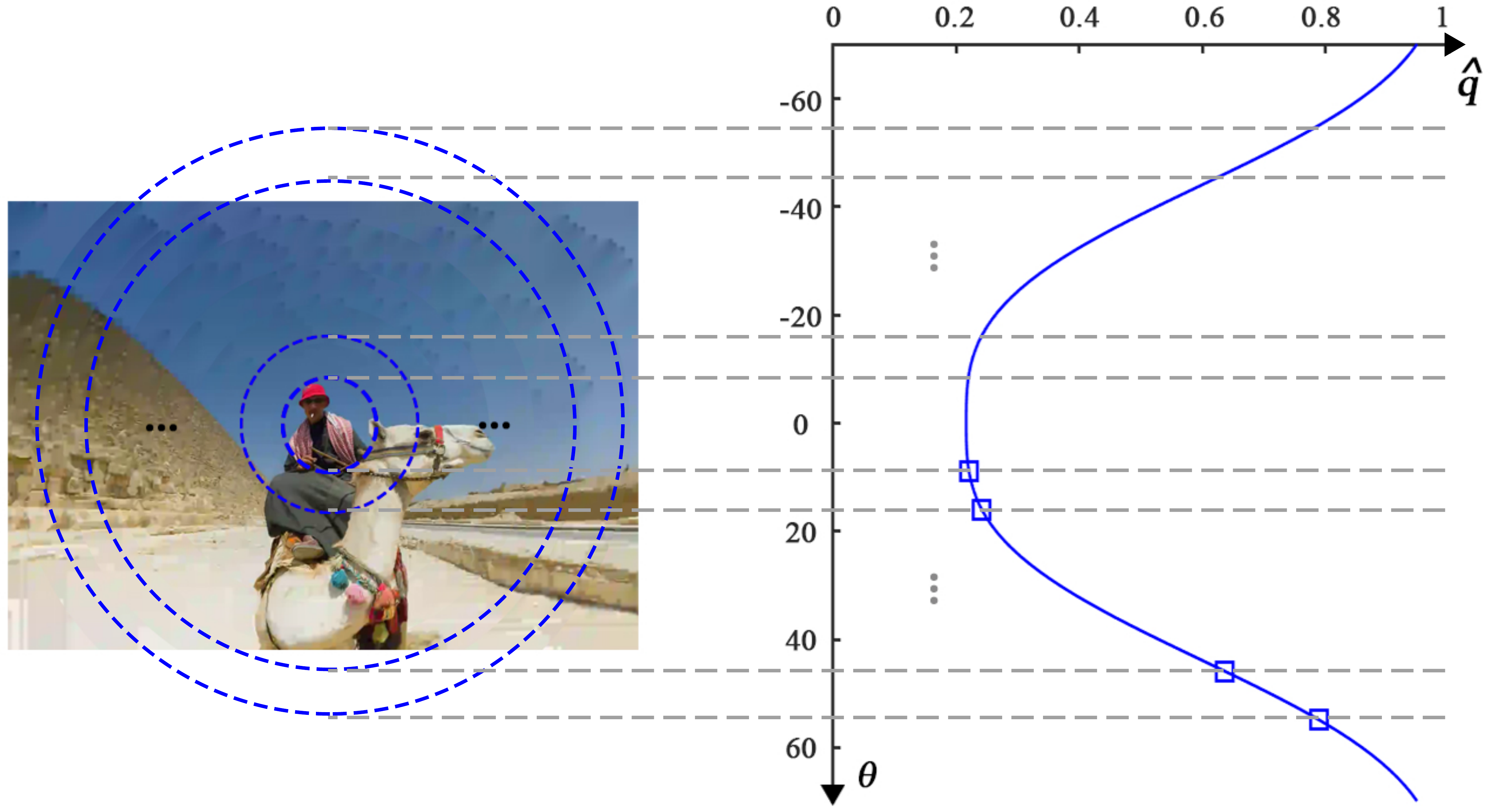}
	\caption{An image with non-uniform quality using various $q$ in central and peripheral areas via model~\eqref{eq:q_s_model}.}
	\label{fig:cross_validation}
\end{figure}

For individual $q$-impact and $s$-impact, we invite another individual subjects to participate the independent cross-validation assessments. Six images (image (a)- (f) shown in Fig.~\ref{fig:sourcedata}) are randomly selected from SUN360 database~\cite{xiao2012recognizing}.
Different from the model development where we combine the double stimulus and JND to find out the impacts of the
quantization and spatial resolution on the perceptual quality with respect to the eccentric angle $\theta$, we
propose to directly measure the MOSs for each image pair.
One is with the uniform quality using $q_c$ or $s_c$ (i.e., for simplicity, we could initially set $q_c$ = 8, $s_c$ = 4096$\times$2160 or other native resolutions); and the other one is with
non-uniform quality using $q(\theta)$ or, $s(\theta)$.
For non-uniform quality, the entire image is divided into seven parts to ensure the smooth quality transition across regions as shown in Fig.~\ref{fig:cross_validation}.  Associated $q$ or $s$ (or corresponding $\hat{q}$ or $\hat{s}$) are derived through developed model in~\eqref{eq:q_s_model}.
Note that except those fixed parameters, content features are extracted from the images to derive the corresponding $c$ via~\eqref{parameter-c-prediction} explicitly.

To let  the participants familiarize themselves with the quality scales from the worst (MOS = 0) to the best (MOS = 10), we prepare the test samples for the {\it Attic} and  {\it Dessert} image in Fig.~\ref{fig:sourcedata_model_development} with uniform  quality at different  scales through multiple $q$s or $s$s. We then mix the test image pairs from all test images shown in Fig.~\ref{fig:sourcedata}, and  place them randomly to collect MOSs.
Subject is asked to give its MOS from 0 to 10 for each displayed sample sequentially.
Note that each image sample repeats three times in total. Thus, the same content is repeated for six times: three for the uniform quality, and other three for the copy with non-uniform quality.  Intuitively, the MOSs for each test sample should be very close from a specific subject.  We enforce the repetition to avoid the random noise.
\begin{figure}[b]
	\centering
	\includegraphics[scale=0.65]{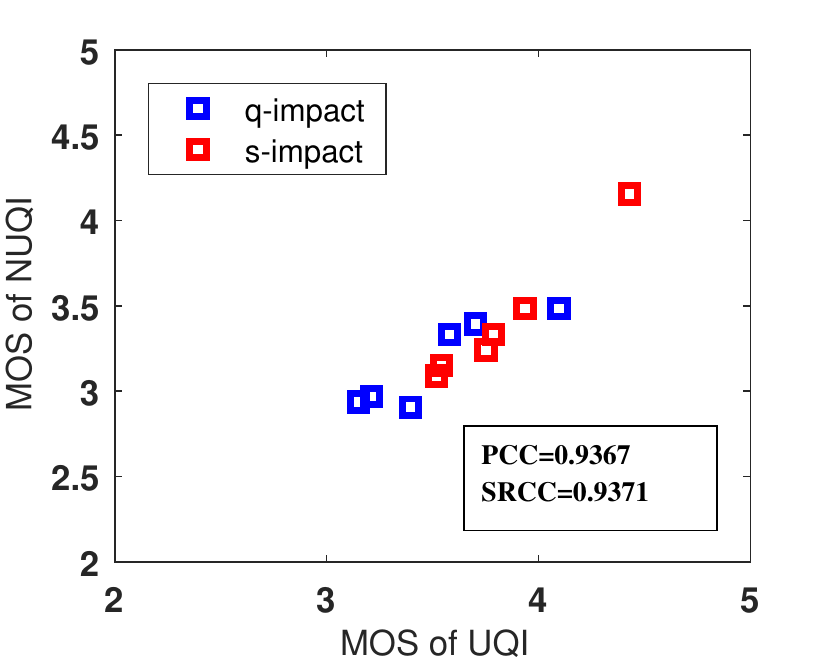}
	\caption{Illustration of measured MOS for the images with uniform quality (UQI) versus corresponding ones with non-uniform quality (NUQI) considering $q$-impact or $s$-impact separately. }
	\label{fig:individual_pcc}
\end{figure}


For each image sample, all raw MOSs from all subjects are collected and then processed following the same screening method discussed in~Sec.~\ref{ssec:subjective_assessment}. Finally, the averaged value is referred to as its MOS. We then plot the MOSs for the samples with uniform quality versus the MOSs for the samples with non-uniform quality, of the same image content, in Fig.~\ref{fig:individual_pcc} for respective $q$-impact and $s$-impact.

We analyze the mean MOS of the image with uniform quality as well as corresponding non-uniform quality. As 
the comparison demonstrates, more than ninety percent of participants cannot  distinguish image sample with uniform or non-uniform
quality configuration. Both high PCC and SRCC in Fig.~\ref{fig:individual_pcc} have shown that the MOSs of non-uniform quality samples are highly correlated with
the MOSs of  uniform quality samples, implying the impressive efficiency of our proposed individual $q(\theta)$ and $s(\theta)$ to model the peripheral vision sensitivity impact quantatitively.

\subsection{Cross-Validation of Joint Impacts of Quantization and Spatial Resolution}
This section extends the cross-validation to the scenarios that impacts of the quantization and spatial resolutions are considered jointly. Quality variation of all twelve images (shown in Fig.~\ref{fig:sourcedata}) are assessed by more than fifty subjects.
In addition to the native spatial resolution at $s_{\max}$, we also prepare another two discrete spatial resolutions at 3584$\times$1890 and 1523$\times$810.
At each $s$,  we compress the different regions with corresponding $q(\theta)$ following the \eqref{eq:q_s_model} (cf. Fig.~\ref{fig:cross_validation}).  We then apply the same procedure as discussed in Sec.~\ref{ssec:x_valid_q_s_impact} to collect the MOSs (i.e., for both uniform and non-uniform quality). Here, subjects rates range from 0 (bad) to 10 (excellent)~\cite{BT500}. All raw MOSs are screened as well to reduce the rating noise.
Figure~\ref{fig:pcc}(a) reveals that most of the subjects can not tell the difference between the uniform and non-uniform quality copies of the same content, with very high PCC at 0.964 and SRCC at 0.953.

\subsection{Self-Adaptive MOS Prediction}
Aforementioned cross-validations mainly compares the image pairs where one is prepared with conventional uniform quality and the
other one is with the non-uniform quality via applying different $q$ and/or $s$ among different visual areas.  This section we explore the
possibility to predict the MOS through the models in~\eqref{eq:s_model},~\eqref{eq:q_model} and~\eqref{eq:q_s_model},
and evaluate the correlations against the collected MOSs via the subjective assessment.

\begin{figure}[t]
	\centering
	\subfigure[]{\includegraphics[scale=0.5]{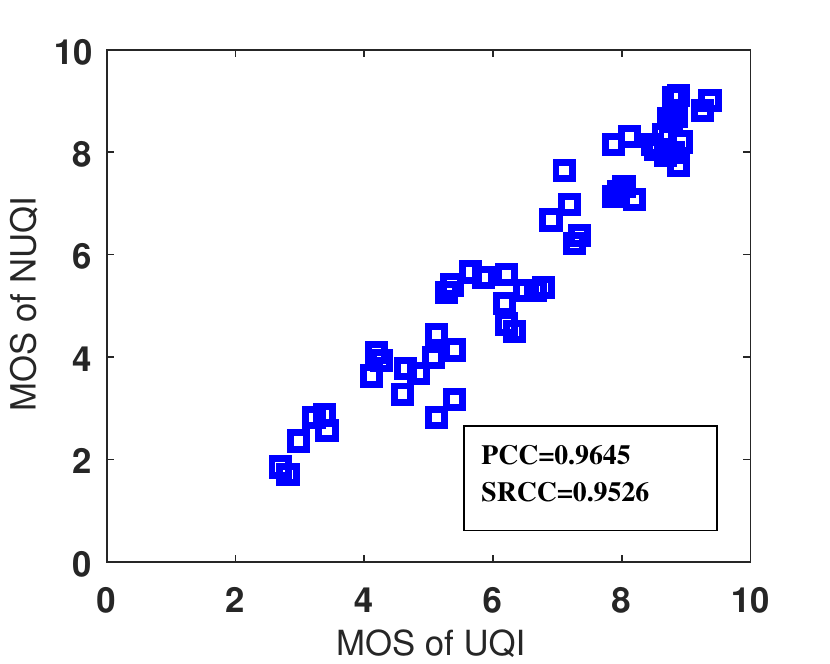}}
    \subfigure[]{\includegraphics[scale=0.5]{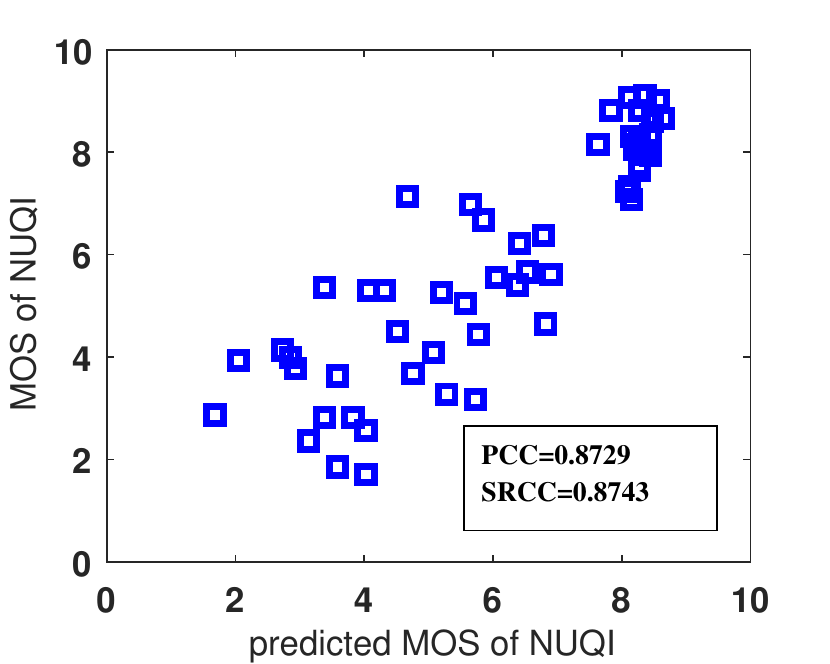}}
	\caption{Illustration of (a) measured MOS of UQI versus its corresponding NUQI (b) predicted MOS of NUQI versus its measured MOS.}
	\label{fig:pcc}
\end{figure}

Since we focus our attention within current FoV when wearing the HMD. Instead of extracting the content features from the entire immersive
image in~\cite{Xiaokai_TIP}, we propose to extract the features within current FoV only to derive the model parameters. With these content adaptive parameters, we could easily estimate the MOS in a straightforward fashion. Figure~\ref{fig:pcc}(b)  illustrates the predicted MOS versus the collected MOS for those image samples prepared with the unequal quality scales among central and peripheral vision areas, resulting in averaged PCC and SRCC more than 0.87.

\section{Application}\label{sec:application}

With the advances in both imaging and display technology, image resolution gets improved significantly, from conventional megapixel scale (such as 1080p, 4K) to the gigapixel scale (such as 40K, 64K)~\cite{david_giga}. With the image rendered in such high pixel resolution, it could cover an ultra wide scene and also provide thumbnail scale texture details simultaneously.  For a typical display with 1080p or 4K resolution in the market, we could zoom in/out to navigate the gigapixel image to locate the region of interest.  To ensure the smooth navigation, we have proposed a multi-scale pipeline and devised our models to enable the ultra-low-latency content retrieval without any loss of the perceptual quality.

\begin{figure}[t]
\centering
\subfigure[]{\includegraphics[scale=0.35]{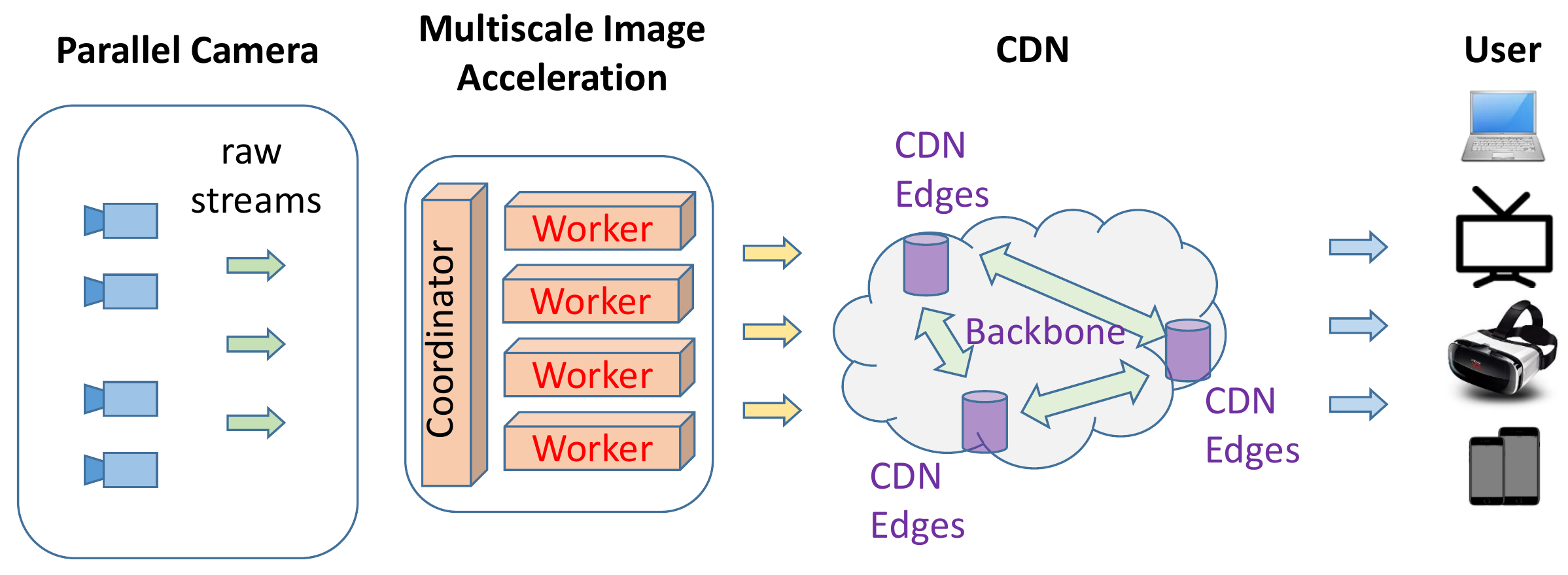} \label{sfig:futureTV_mia}}
\subfigure[]{\includegraphics[scale=0.035]{./Figs/giga/giga_workflowv4.pdf} \label{sfig:giga_transmission}}
\subfigure[]{\includegraphics[scale=0.12]{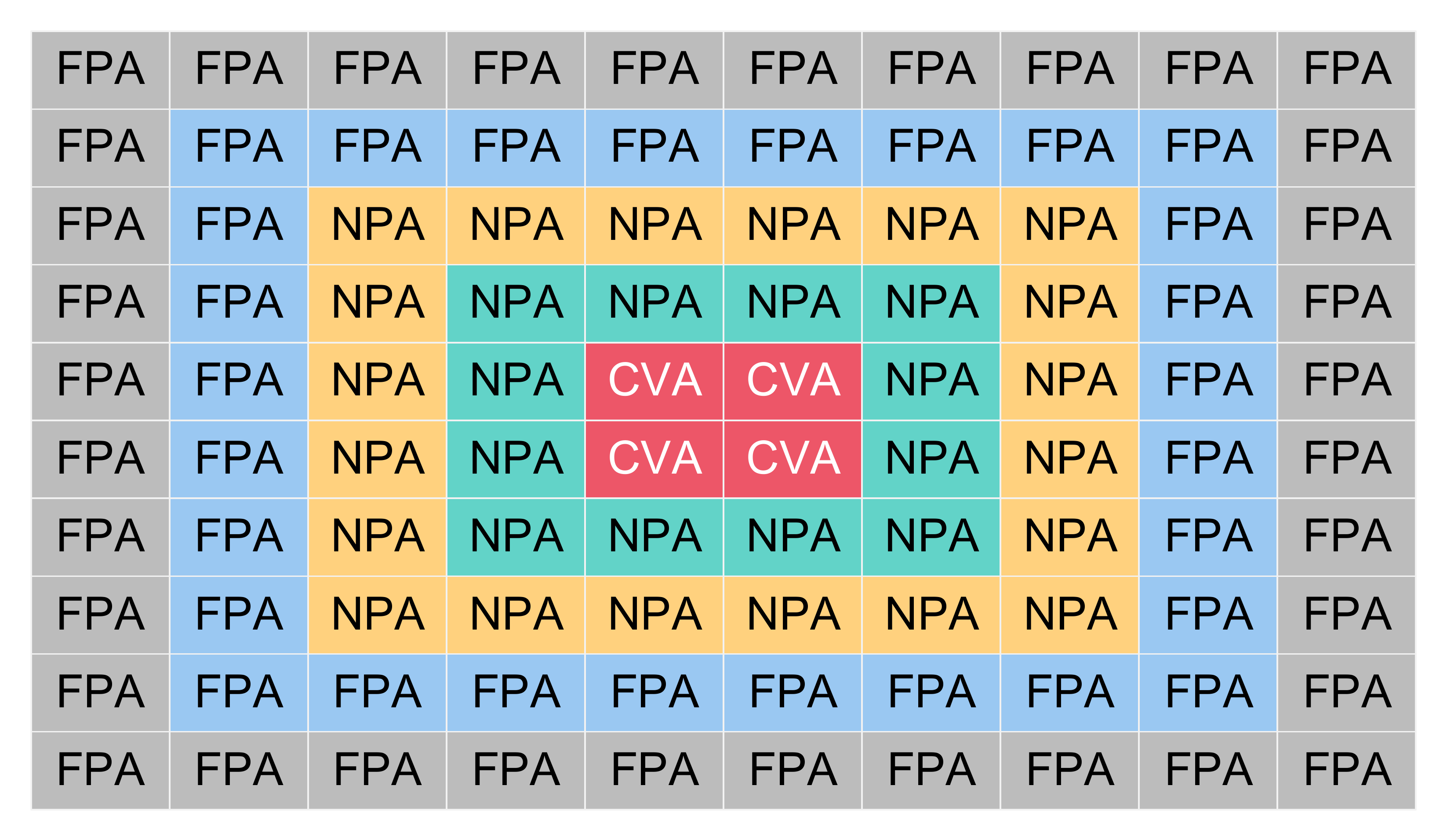} \label{sfig:hmd_fov}}
\subfigure[]{\includegraphics[scale=0.12]{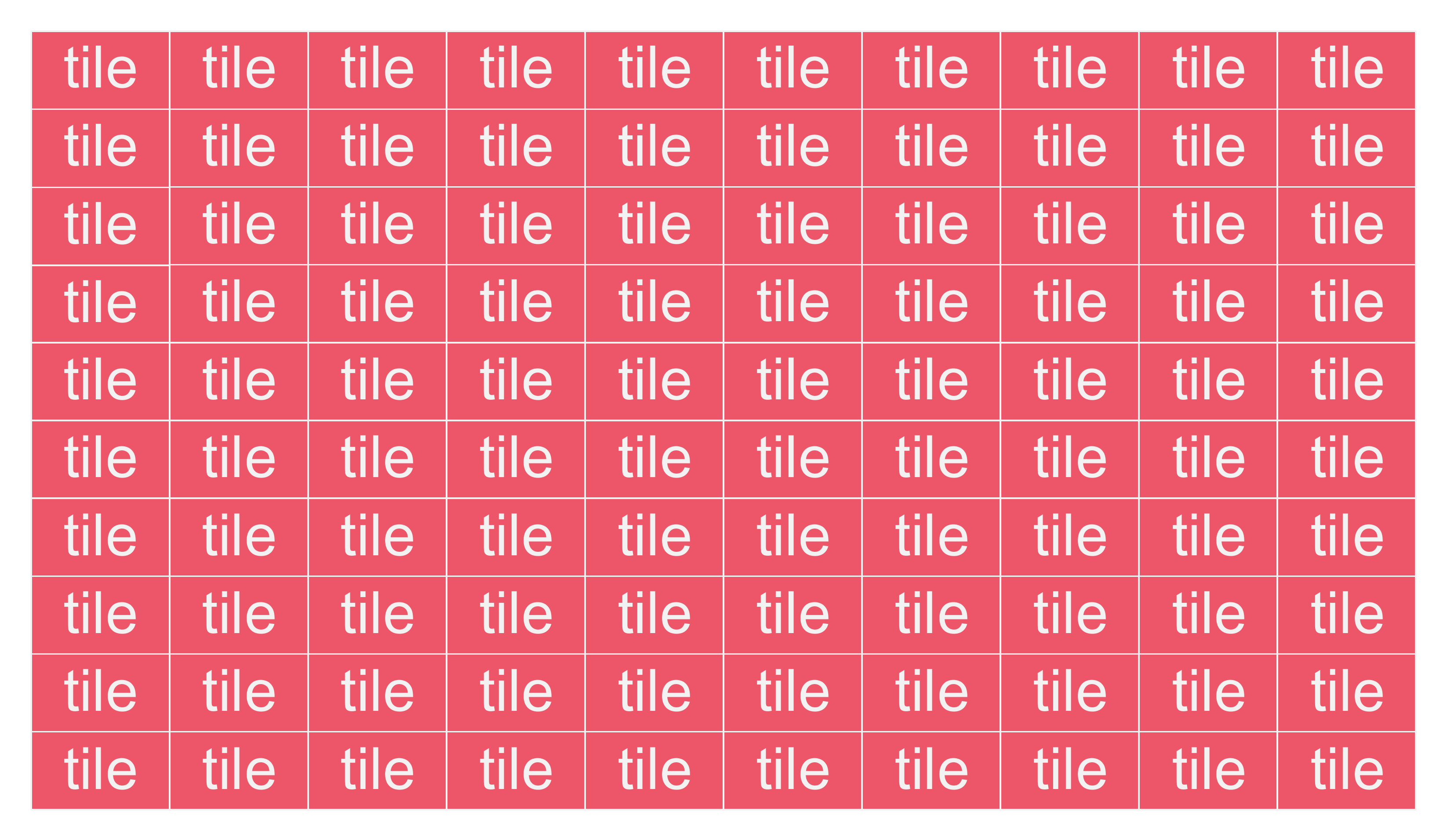} \label{sfig:hmd_fov_uniform}}
\caption{Illustration of a gigapixel streaming system: (a) end-to-end architectural overview (b) multi-scale structure considering the variations from both spatial resolution and quantization  (c) different quality scales applied among distinct vision areas of our FoV with HMD (d) uniform quality applied in our FoV with HMD}
\label{fig:future_tv}
\end{figure}

\subsection{System Architecture}
 Figure~\ref{fig:future_tv} illustrates the end-to-end pipeline from the image capturing to final display.
Such gigapixel image/video can be captured via a parallel or array camera~\cite{parallel_camera}.
One example is the Mantis 70 from Aqueti Inc. (\url{www.aqueti.com}). The Aqueti Mantis 70 camera is an array of 18
narrow field microcameras each with a 25 mm focal length lens
and a 1.6 micron pixel pitch. Each uses a Sony IMX 274 color
CMOS sensor. Sensor read-out, ISP and data compression is implemented
using an array of NVIDIA Tegra TX1 (\url{http://www.nvidia.com/object/tegra-x1-processor.html}) modules with 2 sensors per TX1.
The sensors are arrayed to cover a
73 horizontal field of view and a 21 degree vertical field of view.
The instantaneous field of view is 65 microradians and the fully
stitched image has a native resolution of 107 megapixels.

For each 107 megapixel image sampled at RGB color space, it requires the network bandwidth at (3$\times$107$\times$5)/20 $\approx$  642 Mbps. Here, we assume the compression ratio is 20$\times$ of a typical image coding method (such as the H.264/AVC Intra~\cite{AVC}) and 5 FPS continuous snapshots speed of a mainstream DSLR. Even for the image sampled at YUV420 space, it still demands stable 341 Mbps to sustain the high quality image streaming and rendering over the Internet.  This actually imposes the huge challenge to enable the real-time gigapixel image playback. Thus, we have developed a multi-scale image acceleration (MIA) engine that would process each gigapixel image into multiple quality scales (i.e., via a variety of $q$ and $s$) according to our proposed models in~\eqref{eq:q_s_model}.

The MIA engine enables the massive parallel processing so as to produce the each basic processing unit at multiple quality scales in real-time (or even much faster). Here, the basic processing unit is often referred to as the ``tile''.   As shown  in Fig.~\ref{sfig:futureTV_mia}, the MIA mainly includes two parts: one is the {\it coordinator} that captures the raw inputs from a parallel camera, scales the spatial resolutions and distributes the tiles into multiple parallel {\it workers}; and the other one is the {\it worker} focusing on
 the real-time bitrate transcoding where any high-quality tile video input is transcoded into various versions with multiple quantization applied.
With various combinations of the  spatial resolution and quantization, we can offer a great amount of quality (and bitrate~\cite{Ma_RQModel}) scales to enable the real-time delivery over the existing network. Bit stream at various quality scales (or bit rate scales) of the same content are
used to combat the network dynamics, such as congestion, fading, etc~\cite{DASH}.

In other end, users could wear a Samsung S7 powered GearVR to perform the image navigation in a virtualized space (such as zoom in/out, 6 degree-of-freedom movement, etc), shown in Fig.~\ref{sfig:futureTV_mia} as well. Samsung S7 features a 2560$\times$1440 display, covering
the content of our current FoV in the front.
Typically, a specific FoV consists of one or more tiles (or tile videos).
FoV or viewport navigation can be easily achieved via the tile video adaptation.
In practice, we need to setup an appropriate tile size to fully leverage the peripheral vision model and carefully balance the trade-off between the coding efficiency and processing parallelism. In this work, we set the spatial resolution of a tile at 256$\times$144.

\begin{figure*}[t]
\centering
\subfigure[]{\includegraphics[scale=0.025]{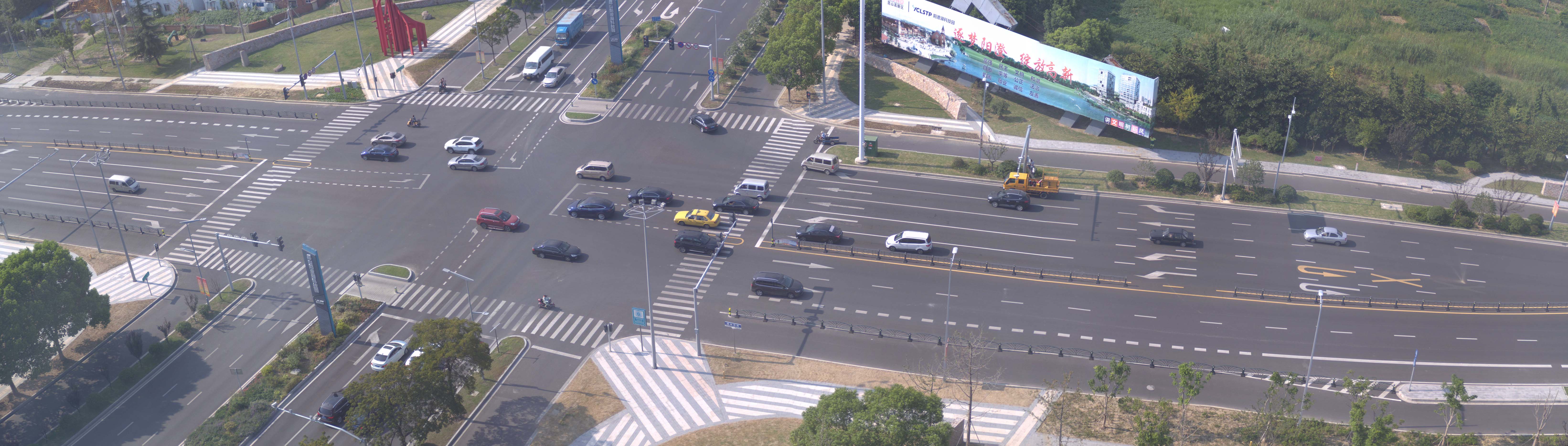} \label{sfig:Road}}
\subfigure[]{\includegraphics[scale=0.025]{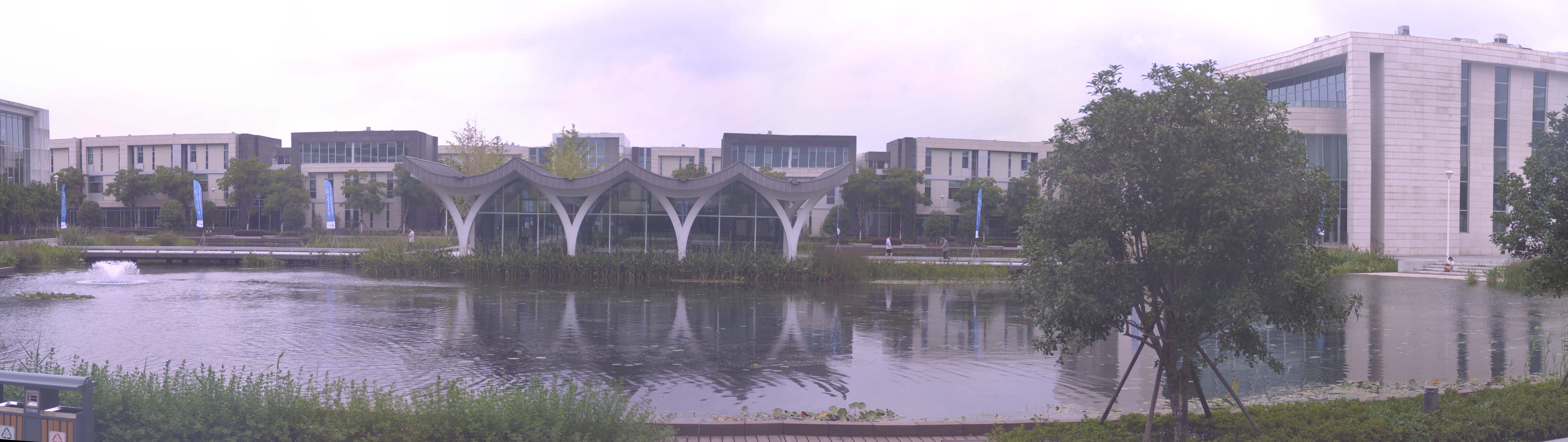} \label{sfig:Park}}
\subfigure[]{\includegraphics[scale=0.025]{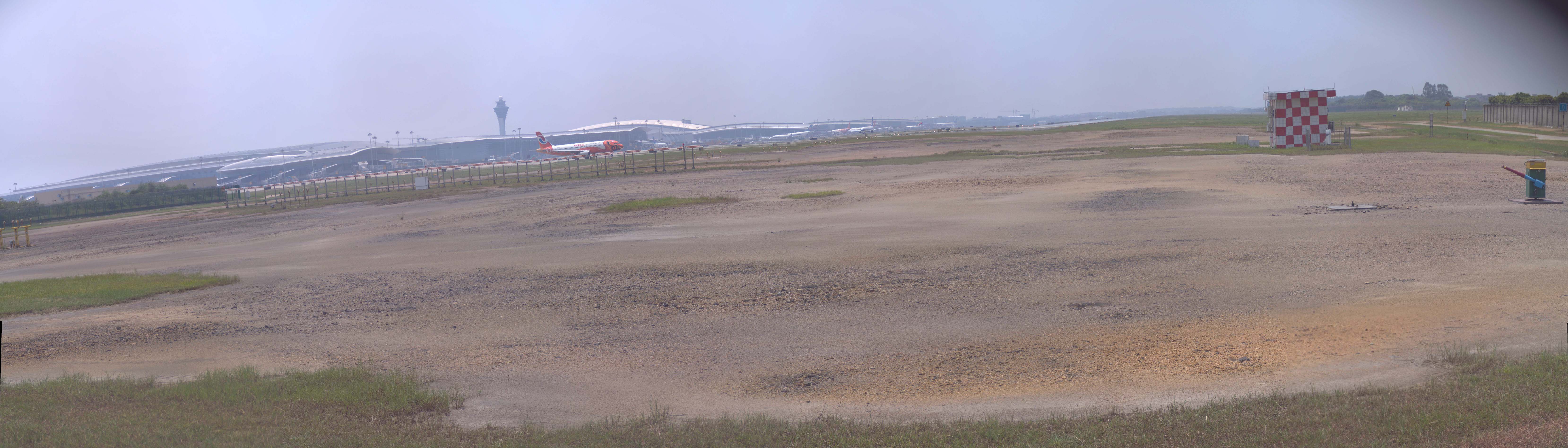} \label{sfig:Airport}}
\caption{Gigapixel source images used for performance comparison: (a) Road (traffic intersection) (b) Park (university campus) (c) Airport}
\label{fig:giga_source}
\end{figure*}

\subsection{Performance Comparison}

\begin{figure}[t]
\centering
\subfigure[]{\includegraphics[scale=0.5]{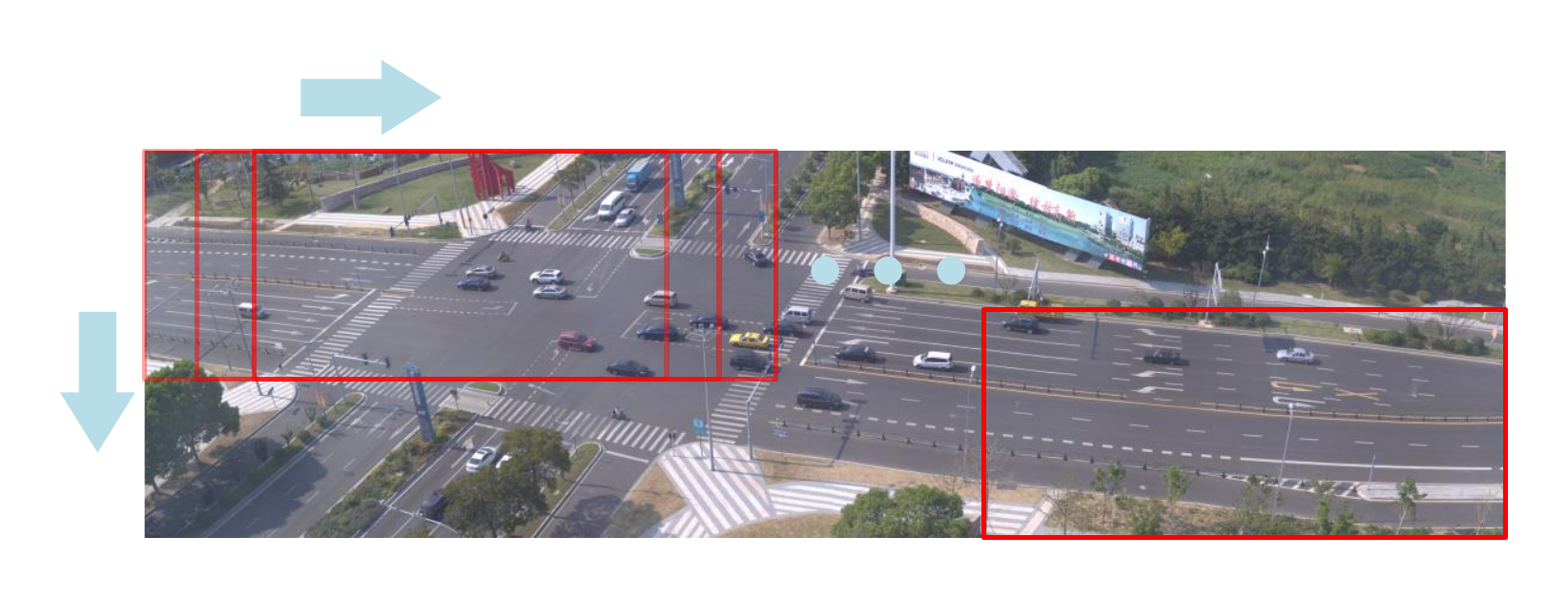}\label{sfig:fov_navi}}
\subfigure[]{\includegraphics[scale=0.8]{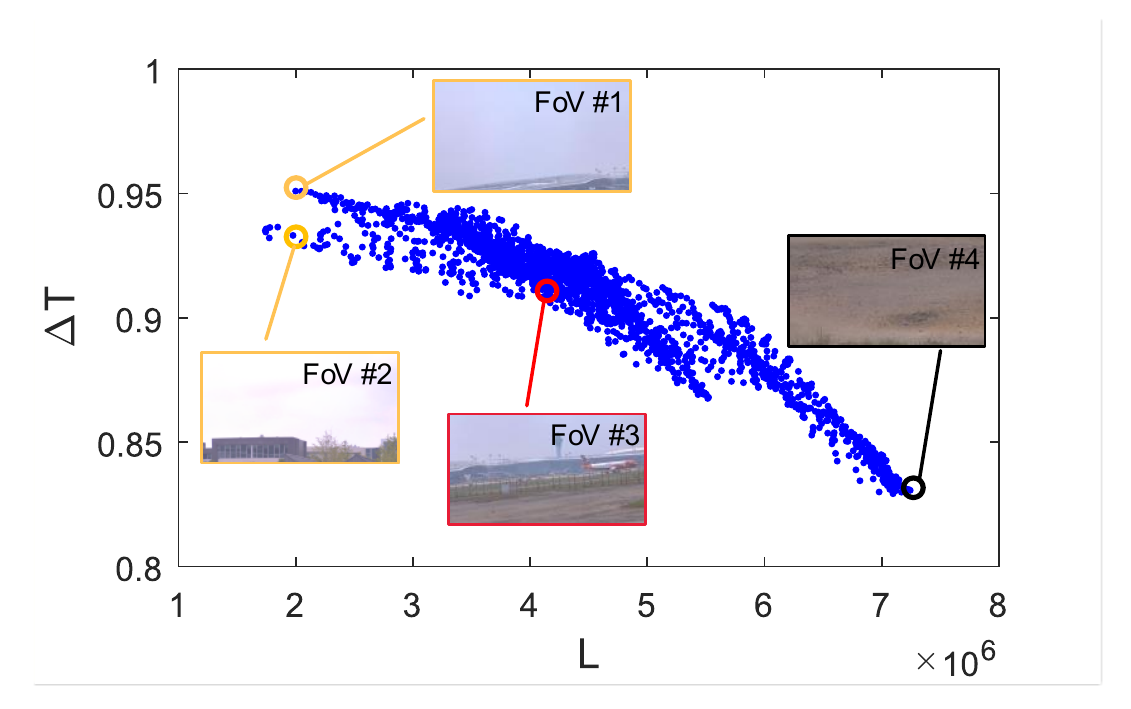}\label{sfig:fov_navi_stat}}
\caption{Illustration of the (a) gigapixel image navigation (b) retrieval time reduction ($\Delta T$) distribution with respect to the lossless image size of current FoV ($L$)}
\label{fig:giga_descr}
\end{figure}

In practice, the landscape overview of the gigapixel image is first displayed in GearVR, where the gigapixel image is scaled to fit the S7 screen as aforementioned. The images are shown in Fig.~\ref{fig:giga_source}. We then can zoom the image to the next higher spatial resolution step by step, to show the fine details of a particular region of interest. We can also perform directional navigation to focus our attention to a new FoV, as presented in Fig.~\ref{sfig:fov_navi}.

In default, a gigapixel image is scaled at various spatial resolutions but with quantization fixed at $q_{\min}$ to ensure the high-quality rendering.
Alternatively, besides the spatial resolution scales, we have processed the content at each spatial resolution into multiple quantization scales, shown in Fig.~\ref{sfig:giga_transmission}. We enforce the tiling for both default and proposed schemes.
 For the FoV rendered in HMD, default solution will have uniform quality everywhere shown in Fig.~\ref{sfig:hmd_fov_uniform}, but our proposed scheme would have the highest quality in CVA, reduced quality in NPA and further reduced quality in FPA, as illustration in Fig.~\ref{sfig:hmd_fov}. As demonstrated previously, quality scales can be achieved by adapting quantizaton $q$ and spation resoltion $s$ across various vision areas. In the meantime, it results in the data size reduction because both quality and bit rate of the image/video content are the function of the $q$ and $s$~\cite{R-STAR,pv_mobileQSTAR}. Given the same access network condition (i.e., bandwidth), the smaller data size, the faster image renders. On average, our scheme could save the image retrieval time about 90.18\% shown in Table~\ref{tab:giga_apply2}.

 It is also noted in Table~\ref{tab:giga_apply2} that retrieval time saving is content dependent. Therefore, we perform the image navigation and record the saving percentage ($\Delta T$) in Fig.~\ref{sfig:fov_navi_stat}, with respect to the size of lossless FoV image (coded with $q$ = 0). As shown, we could even achieve almost 20x speedup (about 95\% reduction of retrieval time) when the FoV content demonstrates the smooth distribution (with small SI) dominantly (sky scene highlighted as FoV\#1, FoV\#2); On the contrary, the saving is reduced when the content is complex with complicate texture (grass land highlighted as FoV\#4).

\begin{table*}[t]
    \renewcommand\arraystretch{1.25}
	\centering
    \begin{threeparttable}
	\caption{Averaged retrieval time of the image shown in current FoV with/without peripheral vision model~\eqref{eq:q_s_model}. }
	\label{tab:giga_apply2}
	\begin{tabular}{|c|c|c|c|c|c|c|c|c|c|c|}
		\hline
        \multirow{2}{*}{Scene}&\multicolumn{2}{|c|}{R\tnote{1} = 5Mbps}&\multicolumn{2}{|c|}{R = 10Mbps}&\multicolumn{2}{|c|}{R = 20Mbps}&\multirow{2}{*}{\tabincell{c}{Saving\\ time} }\\\cline{2-7}
	    & $t_{ori}\tnote{2} /s$& $t_{m}\tnote{3}/s$ & $t_{ori}/s$ & $t_{m}/s$ &$t_{ori}/s$ & $t_{m}/s$&  \\
		\hline
        Road& 6.3312 & 0.5381 & 1.5828 & 0.1345 & 3.1656 & 0.2691 & \textbf{91.71\%} \\

     	\hline
        Park  & 6.2912 & 0.5513 & 1.5728 & 0.1378 & 3.1456 & 0.2757 & \textbf{91.50\%} \\

     	\hline
        Airport& 8.0340 & 0.9388 & 2.0085 & 0.2347 & 4.0170 & 0.4694 & \textbf{89.18\%} \\
        \hline
        average \tnote{4} & 6.8855 & 0.6761 & 1.7214 & 0.1690 & 3.4427 & 0.3381 & \textbf{90.18\%} \\
        \hline
	\end{tabular}
    \begin{tablenotes}
    \item [1] R represents transmission rate.
    \item [2] $t_{ori}$ represents the loading time of a original gigapixel-scale image.
    \item [3] $t_{m}$ represents the loading time of a gigapixel-scale image with non-uniform quality model applied.
    \item [4] In average scene, we calculate the averaged retrieval time among the above all scenes.
    \end{tablenotes}

   \end{threeparttable}
\end{table*}

\section{Conclusion} \label{sec:conclusion}
Non-uniform distribution of photoreceptors on human retina infers that visual perception on natural image is less sensitive in peripheral vision area than in central vision area. We have studied such peripheral vision impact on perceptual quality of immersive images rendered in virtual reality equipped head mounted display, and derived closed-form theoretical models that explicitly describe the control factors (i.e., quantization stepsize, spatial resolution and their joint effect) of image quality with respect to the degree of the eccentricity.

Models are well explained by an unified parametric generalized Gaussian function where parameters are either fixed or can be well estimated by the features extracted from the native content. We randomly select another set of images, extract their features and form the models to guide the non-uniform quality assignment in different regions and predict the subjective quality of the image. Independent cross validation is conducted to demonstrate that users could not tell the difference between uniform and non-uniform quality assignments in peripheral vision. This evident the efficiency of our proposed models.

We further devise our proposed model to apply the non-uniform quality setup for a real-time gigapixel imaging and rendering system.
Compared with the legacy scenario that images are exisited with uniform quality, our model guided non-uniform quality scales among
various tile videos could significantly reduce the image size and therefore improve the rendering throughput about 10$\times$, without noticeable
perceptual quality degradation.

As the future work, we will combine this static spatial vision
study with the temporal variation that is often happened
when users navigate the content inside an virtualized environment. Such navigation induced quality impact exploration is also stuided
in our another work~\cite{shaowei_ISCAS}.  We also would like to make our data public accessible at \url{http://vision.nju.edu.cn/immersive_video/} and encourage more participants from the society to work on this avenue.

\section*{Acknowledgment}
The authors would like to thank all volunteers for their contribution in subjective assessments.

\bibliographystyle{IEEEtran}
\bibliography{2017_TBC_PeripheralVision}

\end{document}